\documentclass[11pt]{article}
\usepackage{latexsym,amsmath,amsfonts,amssymb,amsthm,amsbsy,bm,calrsfs,stmaryrd,color}
\usepackage{hyperref}
\usepackage{xy}
\xyoption{all}
\usepackage{times}

\def\bea{\begin{eqnarray}}
\def\eea{\end{eqnarray}}
\def\be{\begin{equation}}
\def\ee{\end{equation}}
\def\ba{\begin{array}}
\def\ea{\end{array}}

\def\a{\alpha}
\def\b{\beta}
\def\c{\gamma}
\def\d{\delta}
\def\e{\epsilon}
\def\vare{\varepsilon}

\def\l{\lambda}
\def\O{\Omega}

\def\th{\theta}

\def\ad{\dot{\alpha}}

\def\yb{\bar{y}}
\def\zb{\bar{z}}

\def\ad{\dot{\alpha}}

\definecolor{rougef}{rgb}{0.56,0,0}
\definecolor{vertf}{rgb}{0,0.5,0}
\definecolor{bleuf}{rgb}{0,0,0.8}

\begin{document}
\renewcommand{\thefootnote}{\fnsymbol{footnote}}
%

\vspace{7mm}

\begin{center}
{\Large\bf A minimal BV action for Vasiliev's four-dimensional higher spin gravity}
\vspace{1.5cm}

N i c o l a s~~ B o u l a n g e r \footnote{F.R.S.-FNRS Associate Researcher}~, ~~ 
N i c o l \`o ~~ C o l o m b o 
~~~~ and ~~~~P e r ~~ S u n d e l l\footnote{F.R.S.-FNRS Researcher with an 
Ulysse Incentive Grant for Mobility in Scientific Research}

\vspace{15mm}

\textit{Service de M\'ecanique et Gravitation\\Universit\'e de Mons --- UMONS \\
20 Place du Parc\\ B-7000 Mons, Belgium}
\vspace*{1cm}

{\footnotesize \tt nicolas.boulanger@umons.ac.be, nicolo.colombo@umons.ac.be, per.sundell@umons.ac.be}

\end{center}

\vspace{1.5cm}

\begin{minipage}{.90\textwidth}

\textsc{Abstract.}

The action principle for Vasiliev's four-dimensional higher-spin gravity 
proposed recently by two of the authors, 
is converted into a minimal BV master action using the AKSZ procedure, 
which amounts to replacing the classical 
differential forms by vectorial superfields of fixed total degree given by 
the sum of form degree and ghost number. 
The nilpotency of the BRST operator is achieved by imposing 
boundary conditions 
and choosing appropriate gauge transitions between charts 
leading to a globally-defined formulation based on a principal bundle.

\end{minipage}

\vspace{.9cm}

\renewcommand{\thefootnote}{\arabic{footnote}}

\setcounter{footnote}{0}

\newpage

{\small \tableofcontents }

\section{Introduction} \label{sec:Intro}

In \cite{Boulanger:2011dd}, Vasiliev's four-dimensional 
higher-spin gravities \cite{Vasiliev:1990en,Vasiliev:1990vu,Vasiliev:1992av},
including the minimal bosonic models \cite{Sezgin:2002ru},
have been equipped with action principles of generalized-Hamiltonian type. 
The properties of Vasiliev's theory that underlie the construction of the actions 
hold true 
in general models with Lorentzian signature and negative cosmological constant, 
including models with Yang-Mills sectors and supersymmetries.
The off-shell formulation of \cite{Boulanger:2011dd} combine 
the principle of unfolding 
\cite{Vasiliev:1988xc,Vasiliev:1990en,Vasiliev:1988sa,Vasiliev:1992gr,Vasiliev:2005zu}, 
which lies at the heart of Vasiliev's equations,
with a natural extension of the generalized Poisson sigma model
from graded-commutative to graded-associative differential algebras
\footnote{Preliminary investigations indicate a further natural extension 
to homotopy-associative differential algebras.}.

In the graded-commutative case, the generalized Poisson sigma model 
was first studied within the  two-dimensional context 
\cite{Ikeda:1993fh,Schaller:1994es,Cattaneo:2001ys} whose 
Batalin--Vilkovisky 
(BV) formulation \cite{Batalin:1981jr,Batalin:1984jr} was geometrized by 
Alexandrov--Kontsevish--Schwarz--Zaboronski (AKSZ) in 
\cite{Alexandrov:1995kv}, later used for a perturbative path-integral 
derivation \cite{Cattaneo:1999fm,Cattaneo:2001ys,Cattaneo:2001bp} of 
Kontsevish's star-product \cite{Kontsevich:1997vb} on Poisson manifolds. 
These models are closely related to topological BF models 
(see the review \cite{Birmingham:1991ty} and refs. therein); 
interestingly, the BF model without Poisson structure on a non-
commutative manifold was studied in \cite{Blasi:2005vf,Vilar:2007iu}.
Further developments of the AKSZ formalism can be found in 
\cite{Grigoriev:1999qz,Park:2000au,Hofman:2002rv} and 
\cite{Ikeda:2000yq,Ikeda:2001fq,Roytenberg:2002nu,Hofman:2002jz,Ikeda:2002wh,Ikeda:2006wd,Roytenberg:2006qz,Barnich:2009jy,Ikeda:2012pv}, 
and its close ties to unfolded dynamics have been stressed in 
\cite{Barnich:2004cr,Barnich:2005ru,Grigoriev:2006tt,Barnich:2010sw,Grigoriev:2010ic,Kaparulin:2011zz}.
For related treatments of more general dynamical systems, not 
necessarily based on differential algebras, see
\cite{Kazinski:2005eb,Lyakhovich:2006sc} 
and references therein.

The two main results of this paper are:
\begin{itemize}
\item a set of conditions on the couplings in the generalized 
Hamiltonian (see Eq. (\ref{Kconditions}) and (\ref{Pi0conditions})) 
and on the boundary values of certain fields and gauge parameters 
(see Eq. (\ref{Bconditions})), that together assure the existence of a 
globally-defined action principle of fiber-bundle type on a base 
manifold with non-trivial atlas and boundaries;
\item an extension of the AKSZ formalism to unfolded systems on 
non-commutative base manifolds, in such a way as to construct a 
minimal BV-AKSZ master action for Vasiliev's four-dimensional 
higher spin gravities (see Eqs. (\ref{AKSZ1}) and (\ref{AKSZ2})).
\end{itemize}

In all types of generalized Poisson sigma models, whether on commutative 
or non-commutative base manifolds, 
the physical degrees of freedom are contained in boundary vertex 
operators \cite{Cattaneo:2001ys,Park:2000au}.
The boundary lives in a graded target-space manifold equipped 
with a nilpotent vector field of degree $1$, referred to as the 
$Q$-structure, and compatible poly-vector fields of suitable degrees 
depending on the dimension of the base manifold, 
whose mutual Schouten brackets vanish, thus defining a 
generalized Poisson structure referred to as a $QP$-structure 
in the bi-vector case\footnote{which is equivalent to a pure Poisson 
structure by means of a large 
graded canonical transformation that exchanges zero-forms and 
one-forms.}; see \cite{Zucchini:2008hn} and refs. therein.
The bulk lives in a suitably parity-shifted phase space of the boundary target space 
such that each boundary field becomes paired with a momentum in its
turn constrained on the boundary, which thus breaks the group of 
canonical  transformations.  
Assuming a single boundary, the classical limit is thus determined 
by the $Q$-structure and the choice of global formulation used
to construct the boundary observables, \emph{e.g.} formulation on
a fiber bundle with structure group corresponding to the manifest gauge 
symmetries off shell, as we shall discuss more below; for a related 
analysis, see \cite{Kotov:2007nr,Kotov:2010wr}.

In the AKSZ quantization scheme,
the free theory consists only of the kinetic bulk terms, which do not 
depend on the physical vielbein and hence remain well-defined in 
limits where the metric vanishes.
The latter can be gauge-fixed using an auxiliary metric and the physical states can 
be defined by means of a BRST operator \cite{Becchi:1974xu,Becchi:1974md,Becchi:1975nq,Tyutin:1975qk} 
whose existence is guaranteed, at least semi-classically,
by a vectorial supersymmetry that implies that the AKSZ master action 
obeys classical as well as quantum BV master equations, as we shall discuss below.

The unfolded framework may thus provide a bridge between deformation 
quantization and quantum field theories in their metric phases.
The idea is that the latter phase may arise within the former in suitable global 
formulations allowing 
combinations of nontrivial $P$ structures and boundary vertex 
operators depending algebraically on the physical vielbein.
It may then be possible to draw Feynman diagrams, with propagators only 
in the bulk and vertices in both bulk and boundary, describing quantum 
fluctuations for dynamical boundary fields such as scalars, vectors, metrics 
and higher-spin fields in higher-spin gravities in nontrivial metric 
backgrounds, unlike the case of bulk actions without $P$-structures. 
Another intriguing feature of the AKSZ approach 
is the cancellation of all vacuum bubbles 
in flat auxiliary background metrics, which suggests that the 
Poisson sigma model can be summed over bulk topologies, defining a 
third-quantization on top of the second-quantization, that may thus be of 
importance for addressing the vacuum problem in generally covariant 
quantum field theory.

\subsection{Plan of paper}

The plan of the paper is as follows:
In Section \ref{sec:commuting} we review off-shell formulations of 
unfolded systems on commutative base manifold, paying attention to 
global issues that we have not seen being treated elsewhere to the same level
of completeness.
In Section \ref{sec:noncommut} we extend the AKSZ formalism to
unfolded systems on non-commutative base manifolds in such 
a way as to construct a minimal master action for Vasiliev's theory.  
We conclude in Section \ref{sec:conclusions}. The Appendix details
the usage of vector fields and functional derivatives on non-commutative
manifolds.

\section{Action principles for unfolded systems on 
commutative manifolds}\label{sec:commuting}

\subsection{General ideas}

\paragraph{Unfolded dynamics.}

Unfolded dynamics concerns the formulation of field theory in terms of 
differential algebras. 
In their basic setting, referred to as graded-commutative free differential 
algebras, these are sets of differential forms on ordinary commutative 
(super)manifolds that remain invariant under exterior differentiation and 
wedging. 
Their generating elements, denoted by $X^\a$ below, are locally-defined 
forms whose exterior derivatives are completely constrained in a 
Cartan-integrable fashion, amounting to generalized curvature constraints 
written ${\rm d}X^\a+Q^\a(X)\approx 0$ below.

Various moduli spaces, consisting of gauge orbits subject to boundary 
conditions, including transitions between charts in the interior of the base 
manifold,
are then described by different types of classical observables as follows.
As the observables are dual pairings between elements in the differential 
algebra and geometric structures on the base manifold, such as points, 
curves and cycles, 
they possess two key invariance properties: i) invariance on-shell under 
small diffeomorphisms, preserving the geometric structures;
and ii) invariance off-shell under the generalized structure group 
containing a subset of all Cartan gauge symmetries.

We wish to stress that as for the off-shell gauge structure, \emph{i.e.} 
structure group and the off-shell resolutions of the corresponding set of 
observables, there exist multiple, physically inequivalent choices.
This leads to the notion of a large moduli space of an unfolded system 
containing physically distinct phases, such as for example unbroken and 
metric-like phases of a theory of higher-spin gravity.
The analysis of phase transitions thus requires a framework for computing 
partition functions in different ensembles in which the generators of the 
differential algebra play the role of fundamental fields entering directly 
into the path integral measure.

\paragraph{BV-AKSZ implementation.}

Unfolded dynamics, on commutative as well as non-commutative manifolds,  
admits a natural off-shell formulation of the generalized Hamiltonian type:
the bulk action consists of kinetic terms $\sim \int P\wedge dX\,$, 
where thus $X$ and $P$ may have form degrees greater than one, plus a 
Hamiltonian ${\cal H}(X,P)$ containing all interactions; the latter are 
subject to integrability conditions assuring that the gauge symmetries of 
the kinetic terms are deformed smoothly\footnote{As usual, the term 
``smooth'' refers to constancy of the number of gauge parameters.
However, the ``number'' of physical degrees of freedom, as measured by 
classical observables, may change as non-abelian gauge interactions change 
physical-observable conditions abruptly; secondly, the phase-space volume 
elements themselves depend on strengths of couplings, that may induce 
critical phenomena.
In the case of higher-spin gravities, the free fields are characterized by 
point-wise defined Weyl tensors (polarization tensors), while for fully non-
linear solutions, the physical content in the Weyl tensors is captured by 
non-local observables \cite{Colombo:2010fu} such as the eigenvalues of a 
certain Weyl zero-form operator \cite{Iazeolla:2011cb}. In addition, the 
full solution space exhibits an interesting phase structure with critical 
``electric'' fields \cite{Iazeolla:2011cb}. }
into non-abelian gauge symmetries that need not close off-shell.
The generalized curvature constraints arise on boundaries of bulk 
manifolds -- on which the momenta variables vanish -- 
upon extremizing the action,
and the aforementioned ensembles arise upon perturbing the action by various 
generalized Poisson structures coupling to the bulk and topological vertex 
operators inserted at the boundaries, which one may think of as third-
quantized analogs of closed- and open-string states, respectively. 
These key features of the generalized Hamiltonian action can be incorporated 
into quantization schemes based on the BV field-anti-field formalism 
or generalizations thereof,
which also lends itself to topological summation, master-field descriptions 
of (topological) vertex operators ensembles and other ``third-quantized'' 
concepts, which one may view as being defined in this fashion.
Its precise relation to standard ``second-quantized'' amplitudes remains to 
be established, however, though proposals for how these may arise -- in a 
suitable metric phase -- have been made in the case of higher-spin gravity 
\cite{Sezgin:2011hq}.

As found by AKSZ, the BV formalism can be implemented in the generalized-
Hamiltonian case by extending each differential form into a ``vectorial'' 
superfield of fixed total degree given by the sum of form degrees, 
ranging from zero to the top-form degree, 
and ghost numbers belonging to the integers. 
%
This construction manifests the fact that the (canonical) Poisson bracket in 
target (super)space induces the BV anti-bracket on the space of maps.
As a result, substituting each field in the classical action by its 
corresponding superfield and projecting to zero ghost number yields a master 
action obeying the classical BV master equation and reducing to the 
classical action when all anti-fields vanish. 
Moreover, the corresponding BV Laplacian annihilates any local 
super-functional, and hence in particular the AKSZ master action,
which thus obeys the classical as well as the quantum master equations.
The BRST transformations thus remain canonical at the quantum level, and 
hence, in the absence of anomalies, the quantum field theory will possess a 
BRST operator acting as a differential within a suitable homotopy 
associative algebra.

In what follows we shall describe the BV-AKSZ formalism in more detail,
after which we shall adapt it in the next section to the case of 
Vasiliev's 4D higher-spin gravity theory, which is based on a 
graded-noncommutative and associative free differential algebras.

\subsection{Classical sigma model}
  
\paragraph{Classical unfolded dynamics on commutative manifolds.}

At the classical level, an unfolded system on a commutative base manifold $B$ is 
a graded-commutative free differential algebra ${\cal A}$ on $B$.
Decomposing the base manifold into charts, say $B=\bigcup_\xi B_\xi$, the free 
differential algebras decomposes into sub-algebras, say 
${\cal A}=\bigoplus_\xi {\cal A}_\xi$ that are invariant  
under the wedge product and the action of the exterior derivative d.
The generators $\{X^\a_\xi\}$ of ${\cal A}_\xi$ are thus 
differential forms of degrees $p_\a:={\rm deg}(X^\a_\xi)\geqslant 0$, 
defined locally on $B_\xi$ and valued in some finite-dimensional real 
spaces $\Theta_\a$, referred to as types, and obeying
generalized curvature constraints, 
\emph{viz.} 
\begin{eqnarray}
R^\a_\xi~:=~{\rm d}X^\a_\xi+Q^\a(X^\b_\xi)~\approx~0\ ,
\end{eqnarray}
where $Q^\a$ are wedge functions obeying the structure equations 
\begin{eqnarray}
Q^\b \partial_\b Q^\a~\equiv~0\ ,
\end{eqnarray}
with $\partial_\a$ denoting the left-derivative with respect to $X^\a$.
These identities imply generalized Bianchi identities  (the chart index $\xi$ 
will be omitted from now on whenever ambiguity cannot arise)
\bea 
{\rm d}R^\a -R^\b \partial_\b Q^\a~\equiv~0\ ,
\eea
such that the constraints are universally Cartan integrable, 
\emph{i.e.} compatible 
with ${\rm d}^2\equiv 0$ in arbitrary dimensions.
It follows that the generalized curvatures transform into each other under Cartan 
gauge transformations, \emph{viz.}
\begin{eqnarray}
\delta_\e X^\a~:=~{\rm d} \e^\a-\e^\b \partial_\b Q^\a \quad \Rightarrow\quad 
\delta_\e R^\a~=~(-1)^\beta\e^\b R^\c \partial_{\c}\partial_{\b} Q^\a\ ,
\end{eqnarray}
where $\e^\a$ are unconstrained gauge parameters of degrees 
${\rm deg}(\e^\a)=p_\a-1$ (hence $\e^\a\equiv 0$ if $p_\a=0$) 
valued in $\Theta_\a$ and defined on $B_\xi\,$. 
The locally-defined solution spaces consist of gauge orbits 
\begin{eqnarray}
X^\a_{C;\l}~=~\left.G_\l X^\a \right|_{X^\a=C^\a}
\end{eqnarray}
labeled by zero-form integration constants $C^\a=\delta_{p_\a,0} C^\a $ 
obeying $dC^\a=0$ and generated by finite Cartan gauge transformations 
\begin{eqnarray}
G_\l~:=~\exp \left(({\rm d} \l^\b-\l^\c \partial_\c Q^\b(X))\partial_\b\right)\ ,
\end{eqnarray}
where $\l^\a$ are gauge functions of degrees ${\rm deg}(\l^\a)=p_\a-1$ 
(and hence $\l^\a\equiv 0$ if $p_\a=0$).
The locally-defined solution spaces can be glued together into globally-defined 
solution spaces via gauge transitions, \emph{viz.}
\begin{eqnarray}
X^\a_\xi~=~\left.G_{t^{\xi'}_\xi} X^\a\right|_{X^\a=X^\a_{\xi'}}\ ,
\end{eqnarray}
using suitable locally-defined parameters $t^{\a,\xi'}_\xi$ 
defined on the overlaps $B_\xi\cap B_{\xi'}$.
The choice of the structure group leaves room for various physically distinct
possibilities depending on the $Q$-structure
(for a discussion, see e.g. \cite{Boulanger:2011dd,Sezgin:2011hq}). 
In particular, one may seek global formulations that are direct 
generalizations of fiber bundles with classical observables that are invariant
off-shell under gauge transformations with parameters belonging to the structure 
algebra, and on-shell under the complete Cartan gauge algebra. 
For more general geometric formulations, 
see e.g. \cite{Kotov:2007nr,Kotov:2010wr}.

\paragraph{Classical generalized Hamiltonian action.}

Classical unfolded systems can be embedded into on-shell configurations of 
generalized Poisson sigma models, namely as boundary configurations in 
formulations on open base manifolds of fixed dimension. 
To this end, one assumes that  
\bea 
{\rm dim}(B)~=~\hat p+1\ ,\qquad \partial B~=~\cup_\l B'_\l\ ,
\eea
where each $B'_\l$ is a connected boundary component (which may in itself be 
covered by an atlas inherited from the bulk), and considers sigma-model maps
\begin{eqnarray}
\phi~:~T[1]B~\rightarrow~M
\end{eqnarray}
of vanishing intrinsic degree from the parity-shifted tangent bundle $T[1]B$ 
to a target space $M$ given by an $\mathbb N$-graded symplectic $Q$-manifold. 
The latter consists of charts, 
\begin{align}
M=\cup_I M_{I}\;,
\end{align}
with locally-defined coordinates 
\bea 
Z^i_I~:~M_{I}~\rightarrow~\theta^i[p_i]\ ,\qquad {\rm deg}
(Z^i_I)~=~p_i~\in~\mathbb N\ ,
\eea
where $\theta^i[p_i]$ denote $p_i$-suspended types.
It carries two compatible geometric structures: a symplectic two-form ${\cal O}$ 
of degree $\hat p+2$ and a Hamiltonian function ${\cal H}$ 
of degree $\hat p+1$ obeying the \emph{structure equation}
\begin{eqnarray}
\{{\cal H},{\cal H}\}_{[-\hat p]}&\equiv &0\ ,\qquad 
     {\rm deg}({\cal H}) ~=~\hat p+1\ .
\end{eqnarray}
The canonical Poisson bracket, which has intrinsic degree $-\hat p$ and is graded 
in such a way that $\{{\cal H},{\cal H}\}_{[-\hat p]}$ does not vanish trivially, is 
given by
\begin{equation}
 \{ A,B\}_{[-\hat p]} ~=~ (-1)^{\hat p + (\hat p + i + 1)A} \;\,\partial_iA \; 
 {\cal P}^{ik}\, \partial_jB
\label{PoissonBracket}
\end{equation}
where we use the conventions
\bea 
{\cal O}~=~\tfrac{1}{2}\, {d}Z^i {d}Z^j \widetilde{\cal O}_{ij}
~\equiv~\tfrac{1}{2}\, {d}Z^i{\cal O}_{ij}\,{d}Z^j\ ,\qquad  
{\cal P}^{ik}{\cal O}_{kj}~=~(-1)^{\hat p} \delta^i_j\ .
\eea
In particular, the structure equation reads
\begin{eqnarray}
 (-1)^{i(\hat p+1)}\,
\partial_i{\cal H}{\cal P}^{ij} \partial_j{\cal H}~ \equiv ~0\ .
\label{structure}
\end{eqnarray}
Locally in target space, one can introduce pre-symplectic forms 
\bea 
{\cal O}|_{M_{I}}~=~{\rm d}\vartheta_I\ ,\qquad 
{\rm deg}(\vartheta_I)~=~\hat p+1\ , 
\eea
defined modulo $\vartheta\sim\vartheta+{\rm d}{\cal E}$, and consider the 
generalized Hamiltonian bulk action 
\begin{eqnarray}
S_{\rm bulk}^{\rm cl}[\phi|B]&=& \sum_\xi \int_{B_\xi} {\cal L}_{\xi}^{\rm cl} ~=~  
\sum_\xi \int_{B_\xi} \pi \,\phi^\ast_\xi (\vartheta-{\cal H})\ ,
\end{eqnarray}
where $\phi_\xi\equiv\phi|_{B_\xi}$ and $\pi:\O(T[1]B)\rightarrow\O(B) $ is a degree-preserving canonical homomorphism that takes $k$-forms on $T[1]B$ of degree $p$ to $p$-forms on $B$, \emph{viz.}
\begin{eqnarray}
\pi~:~\Omega^{[k|p]}(T[1]B)\rightarrow \Omega^{[p]}(B)\ ,\eea
and that intertwines the actions of the exterior derivative d in $\O(B)$ and the Lie derivative ${\cal L}_q=i_q\circ {\rm d}-{\rm d} \circ i_q$ in $\O(T[1]B)$ along the canonical $Q$-structure on $T[1]B$ as follows:
\begin{eqnarray}
{\rm d}\circ \pi~=~\pi\circ {\rm d}~=~\pi\circ{\cal L}_q\;, 
 \quad q~:=~\theta^\mu \partial_\mu \; . 
\end{eqnarray}
Equipping $T[1]B$ with coordinates 
\begin{eqnarray}
(x^\mu,\theta^\mu)\ ,\qquad  {\rm deg}(x^\mu,\theta^\mu)~=~(0,1)\ ,
\end{eqnarray}
one has
\bea \pi(f(x^\mu,\th^\mu;dx^\mu,d\theta^\mu))~=~ 
f(x^\mu,dx^\mu;dx^\mu,0) \ .
\eea
Thus the exterior differential d$\,$, which has form-degree one, has degree one, 
\emph{i.e.}
\begin{eqnarray}
{\rm deg}({\rm d})~=~{\rm deg}(q)~=~1\ . 
\end{eqnarray}
The assumption that the sigma-model maps $\phi$ have vanishing intrinsic 
degree implies
\begin{equation}
\Omega^{[k|p]}(M)~\stackrel{\phi^\ast}{\rightarrow} ~\Omega^{[k|p]}(T[1]B)~\stackrel{\pi}{\rightarrow} 
~\O^{[p]}(B)\ ,
\end{equation} 
that is, the pull-back $\phi^\ast$ of a $k$-form of $\mathbb{N}$-degree $p$ on $M$ is a ditto on $T[1]B$, in its turn sent by $\pi$ 
to a $p$-form on $B$; the condition that $M$ is $\mathbb N$-graded (instead of $\mathbb Z$-graded) and ${\rm deg}({\rm d})=1$ implies that 
$p\geqslant k\,$.
Thus, since
\bea {\cal O}~=~{\rm d}\vartheta\in\Omega^{[2|\hat p+2]}(M)\;, \quad 
\vartheta\in \Omega^{[1|\hat p+1]}(M)\;,\quad {\cal H}\in \Omega^{[0|\hat p+1]}(M)\ ,\eea
it follows that 
\bea \pi \phi^\ast (\vartheta-{\cal H})~\in~ \Omega^{[\hat p+1]}(B)\ ,\eea
which can then be integrated by decomposing $B$ into charts $B_\xi$.

\paragraph{Total variation and gauge variation.}

The total variation of the action can be obtained from the Lie derivative 
\begin{equation}
 \{{\rm d},i_{\overrightarrow{\delta Z}}\}({\vartheta}-{\cal H})
~=~i_{{\overrightarrow{\delta Z}}}({\cal O}-{\rm d}{\cal H}) +
 {\rm d}(i_{{\overrightarrow{\delta Z}}}{\vartheta})\ ,
\end{equation}
where the target space vector field 
$\overrightarrow{{\delta Z}}=\delta Z^i\,\vec{\partial}_i\,$.
Writing  
\bea {\vartheta}~= ~{\rm d}Z^i{\vartheta}_i\ ,\eea
one has
\begin{eqnarray}
\delta{\cal L}^{\rm cl}_{\rm bulk}&=&\delta Z^i {\cal R}^j\widetilde{\cal O}_{ij}+
{\rm d}\left(\delta Z^i {\vartheta}_i\right)
\ ,
\label{totalvaria}
\end{eqnarray}
with generalized curvatures and Hamiltonian vector field 
$\overrightarrow{\cal Q}$ defined by
\begin{eqnarray}
 {\cal R}^i &=& {\rm d}Z^i+{\cal Q}^i\ ,
 \qquad {\cal Q}^i\ =\ (-1)^{\hat p+1}{\cal P}^{ij}\partial_j {\cal H} \ ,
\nonumber \\
\overrightarrow{\cal Q} &=& {\cal Q}^i\,\vec{\partial}_i\ ,\qquad {\rm deg}
(\overrightarrow{{\cal Q}})~=~1\ .
\label{Qstruct}
\end{eqnarray}
Demanding the generalized Bianchi identities 
\bea 
{\rm d}{\cal R}^i-{\cal R}^j \partial_j {\cal Q}^i~\equiv ~0\ ,
\eea
requires $\overrightarrow{\cal Q}$ to be a Hamiltonian $Q$-structure, 
\emph{viz.}
\begin{eqnarray}
{\cal L}_{\overrightarrow{\cal Q}}\overrightarrow{\cal Q} &=& \{\overrightarrow{\cal Q},\overrightarrow{\cal Q}\}_{\rm S.B.}~
\equiv~ 0\quad \Leftrightarrow\quad {\cal Q}^j\partial_j {\cal Q}^i~\equiv~0
\quad\Leftrightarrow\quad \partial_i\{{\cal H},{\cal H}\}_{[-\hat p]}\equiv 0\ ,
\end{eqnarray}
which is equivalent to the structure equation assuming there are no constants of 
total degree $\hat p+2\,$. 
The structure equation also implies
\bea {\rm d}(\vartheta -{\cal H})
  ~\equiv~\tfrac{1}{2}\, {\cal R}^i{\cal R}^j\widetilde {\cal O}_{ij}
  \equiv~\tfrac{1}{2}\, {\cal R}^i {\cal O}_{ij}{\cal R}^j\ .\eea
Under the chart-wise defined Cartan gauge transformations
\begin{eqnarray}
\delta_\varepsilon Z^i &:=&
{\rm d} \varepsilon^i -\varepsilon^j\partial_j {\cal Q}^i
+ \tfrac{1}{2}\,\varepsilon^{k}{\cal R}^l \,
\partial_{l}\widetilde{\cal O}_{kj}\,{\cal P}^{ji}\ ,
\end{eqnarray}
the Lagrangian transforms into a total derivative as follows:
\begin{eqnarray} 
\delta_\varepsilon {\cal L}^{\rm cl}_{\rm bulk} &\equiv&
{\rm d}K_\varepsilon\ ,\qquad K_\varepsilon 
~:=~ \varepsilon^i {\cal R}^j\widetilde{\cal O}_{ij}+\delta_\varepsilon
Z^i{\vartheta}_{i}+{\rm d}\Upsilon_\e\ ,
\end{eqnarray}
where $\Upsilon_\e$ is defined on $B_\xi$ and the cancellation of 
${\cal R}^k$-terms requires that
${\cal Q}_i:={\cal O}_{ij}{\cal Q}^j\,$ 
obeys $\partial_i{\cal Q}_j\equiv(-1)^{ij}\partial_j {\cal Q}_i$ 
which holds as a consequence of $\,$d${}^2{\cal H}\equiv0\,$. 
The gauge transformations close as follows \cite{Boulanger:2011dd}:
\begin{eqnarray}
 [\delta_{\varepsilon_1},\delta_{\varepsilon_2}] Z^i &\equiv&
 \delta_{\varepsilon_{12}} Z^i 
 - \overrightarrow{\cal R}\varepsilon_{12}^i
 \ ,\label{closure1}\eea
\bea \varepsilon_{12}^i~\equiv~[\varepsilon_1,\varepsilon_2]^i~:=~ 
-\overrightarrow\varepsilon_{1}\,\overrightarrow\varepsilon_{2}\,
{\cal Q}^i
 ~\equiv~ \overrightarrow\varepsilon_{2}\,\overrightarrow\varepsilon_{1}
\,{\cal Q}^i\ ,
\qquad \overrightarrow{\cal R}~\equiv ~ {\cal R}^i\partial_i\ , 
\label{closure}
\end{eqnarray}
where $\overrightarrow{\cal R}\varepsilon_{12}^i$ generates a trivial gauge transformation $\delta_{\overrightarrow{\cal R}\varepsilon_{12}}$ as can be seen from
\begin{equation}
\delta_{\overrightarrow{\cal R}\varepsilon_{12}} {\cal L}^{\rm cl}_{\rm bulk}(p')~\equiv ~ \int_{p\in B} (\overrightarrow{\cal R}\varepsilon_{12}^i)(p) 
\frac{\delta {\cal L}(p')}{\delta Z^i(p)} ~\equiv~ {\rm d}\left[(\overrightarrow{\cal R}\varepsilon_{12}^i) \vartheta_i
\right](p') \ ,
\end{equation}
which follows from (\ref{totalvaria}).

\paragraph{Global base-manifold formulation of fiber-bundle type.}

The action is well-defined, \emph{i.e.}
\bea 
\delta_\varepsilon S^{\rm cl}_{\rm bulk}~\equiv~ \sum_{\xi} \oint_{\partial B_\xi} K_{\e_\xi}~=~0\ ,
\eea
provided that the locally-defined fields $Z^i_\xi$ and gauge parameters 
$\e^i_\xi$ are subject to suitable conditions at $\partial B_\xi$ -- and we note 
that $\oint_{\partial B_\xi} {\rm d}\Upsilon_{\e_\xi}=0$ since $\Upsilon_{\e_\xi}$ 
is defined on $B_\xi$ and hence globally on $\partial B_\xi$.
Under certain extra assumptions\footnote{For a more general treatment, 
based on geometrical concepts beyond those of the standard theory of fiber 
bundles which are used in the present paper, see 
\cite{Kotov:2007nr,Kotov:2010wr}.} 
on $\vartheta$ and ${\cal H}\,$, the latter amount to conditions at $\partial B$ 
together with rules for gauge transitions $\hat\delta_{t_{\xi'}^{\xi}}$ across chart 
boundaries 
with parameters $t^{i,\xi}_{\xi'}\,$ defined on overlaps. The assumptions are
\begin{eqnarray}
(i) \quad \hat\delta_t K_\varepsilon &\equiv& 0\ ,\quad (ii)\quad 
\partial_j\partial_k 
\overrightarrow{t}{\cal Q}^i~\equiv~0\ ,\quad (iii) \quad K_t~\equiv~0  \ .
\label{Kconditions}
\end{eqnarray}
Assumption (i), which states that $K_\varepsilon$ is defined globally, implies the 
cancellation of contributions to $\delta_\varepsilon S^{\rm cl}_{\rm bulk}$ from 
chart boundaries in the interior of the bulk manifold, leaving 
\bea \left.K_{\varepsilon}\right|_{\partial B}~\equiv~0\ ,\eea
as conditions on fields and gauge parameters off shell.
Assumptions (ii) and (iii) ensure compatibility between having, on the one hand, 
gauge transformations 
$\hat \delta_{\varepsilon_\xi}$ on charts acting on fields $Z^i_\xi$ 
and gauge transition parameters $t^{i,\xi}_{\xi'}$ and, on the other hand, gauge 
transitions $\hat\delta_{t^{\xi}_{\xi'}}$ 
between adjacent charts acting acting on fields $Z^i_\xi$ and gauge 
parameters $\varepsilon_\xi\,$. 
As for (ii), the commutativity of the diagram in Figure \ref{figure1} requires
\begin{figure}
\hspace*{3cm}
\xymatrix @!=.3cm
{
&& Z^i_\xi   \ar[rrrr]^{\delta_{\varepsilon_\xi}} \ar[dd]^{\hat\delta_{t}} & & & & 
Z^i_\xi + \delta_{\varepsilon_\xi}Z^i_\xi \ar[dd]^{\hat\delta_{\tilde{t}}} \\
&&                               & & & & \\
&& Z^i_\xi + \delta_{{t^{\xi}_{\xi'}}}Z^i_\xi \ar[rrrr]^{\delta_{\varepsilon_{\xi'}}} 
& & & &
(\ref{compa1})~ {\mbox{and}}~ (\ref{compa2}) 
}
\caption{\textit{Compatibility condition for the fiber bundle \label{figure1}}}
\end{figure}
\begin{eqnarray}
Z^i_\xi + \delta_{\varepsilon_\xi}Z^i_\xi + 
\hat{\delta}_{\tilde{t}^\xi_{\xi'}}(Z^i_\xi + \delta_{\varepsilon_\xi}Z^i_\xi) 
~=~ Z^i_\xi + \delta_{{t^\xi_{\xi'}}}Z^i_\xi + \delta_{\varepsilon_{\xi'}}
(Z^i_\xi +\delta_{{t^\xi_{\xi'}}}Z^i_\xi )\ ,
\label{compa1}
\end{eqnarray}
where $\delta_{\varepsilon_{\xi'}}$ only acts on fields and 
\begin{eqnarray}
{\tilde{t}}{}^\xi_{\,\xi'}~ :=~ {t}^\xi_{\xi'} + 
\hat\delta_{\varepsilon_\xi}{t}^\xi_{\xi'}\ .
\label{compa2}
\end{eqnarray}
As $\hat\delta_{t_{\xi'}^\xi}\varepsilon_\xi$ drops out, the above condition is 
equivalent to 
\bea \delta_{\hat\delta_{\varepsilon_\xi}t^\xi_{\xi'}} 
Z_\xi^i~=~\left(\delta_{t^\xi_{\xi'}}\delta_{\varepsilon_{\xi'}}
   -\delta_{\varepsilon_\xi}\delta_{t^{\xi}_{\xi'}}\right)Z^i_\xi\ ,
\eea
whose consistency requires (ii) and one identifies
\begin{eqnarray}
\hat\delta_{\varepsilon_\xi}{t}^{\xi}_{\xi'}
~=~[ {t}^\xi_{\xi'},\varepsilon_\xi]\ . 
\end{eqnarray}
The transformation $\hat\delta_{t_{\xi'}^\xi}\varepsilon_\xi$ is instead fixed by 
the third assumption \emph{(iii)} which 
ensures the commutativity between (i) and 
$\delta_\varepsilon{\cal L} \equiv$d$K_\varepsilon\,$;
acting with $\hat\delta_t$ on the latter identity using 
$\hat\delta_t \delta_\varepsilon{\cal L}\equiv\delta_t \delta_\varepsilon{\cal L}+ \delta_{\hat\delta_t\varepsilon}{\cal L}$ and (ii) yields
\begin{eqnarray}
{\rm d}\left(K_{\hat\delta_t\varepsilon+[t,\varepsilon]}
+\delta_\varepsilon K_t\right)~=~0
\end{eqnarray}
from which one deduces that
\begin{eqnarray}
\hat\delta_{{t}^\xi_{\xi'}}\varepsilon_\xi ~=~ [ \varepsilon_\xi, {t}^\xi_{\xi'}]
~\equiv~- \hat\delta_{\varepsilon_\xi}{t}_\xi^{\xi'}
\end{eqnarray}
provided that (iii) holds.

\paragraph{Equations of motion} Applying the variational principle to the 
action yields the following equations of motion and boundary conditions:
\bea 
{\cal R}^i~\approx~0\ ,\qquad \left.\delta Z^i 
\vartheta_i\right|_{\partial B}~\approx~0\ .
\eea
We recall that $\left.K_\varepsilon\right|_{\partial B}\equiv 0$ holds off shell as to 
assure the gauge-invariance of the action and hence the gauge-covariance of the 
above equations of motion as well as 
the cancellation of boundary terms in the interior of $B$ in $\delta S$, \emph{i.e.} 
\bea 
\delta_t (\delta Z^i \vartheta_i)~\equiv~ 0\ .
\label{deltatdeltaZ}
\eea

\paragraph{Canonical coordinates.}

We assume\footnote{This assumption implies no loss of generality provided the 
starting point is the classical unfolded systems on $\partial B$ (with target space 
$N$). It does lead to restrictions, however, starting from systems on $B$ (with 
target space $M$) where it excludes models with $\hat p=2(2n+1)$ and 
coordinates in $Z^{i'}$ of degree $p_{i'}=2n+1$ contributing 
$\frac12 dZ^{i'} dZ^{j'} k_{i'j'}$ to ${\cal O}$ where $k_{i'j'}$ is positive definite, 
such as three-dimensional Chern--Simons theories with compact gauge algebra 
$\mathfrak g_k$. 
The latter instead admit formulations as four-dimensional BF-models with action 
$\int_{B}{\rm tr}(T F-\frac1{2k}T^2)$ where $F:={\rm d}A+A^2$\,, which is locally 
on-shell equivalent to 
$\frac k2\oint_{\partial B}{\rm tr }(A\,{\rm d}A+\frac23 A^3)$\,. 
On the other hand, certain non-compact cases admit formulations 
as three-dimensional BF-models. 
For example, for the gauge algebra $\mathfrak g_k \oplus \mathfrak g_{-k}$, 
which is of relevance for three-dimensional vacuum higher-spin gravities, 
one has 
$\frac{k}2 \int_{B}{\rm tr}(A\,{\rm d}A+\frac23 A^2-\widetilde A d\widetilde A-\widetilde A^3+{\rm d}(A\widetilde A))\equiv k \int_B{\rm tr} (E R+\frac1{12} E^3)$ 
where now ${\rm dim}( B)=3$ and $E:=A-\widetilde A$, 
$R:={\rm d}\Gamma+\Gamma^2$ and $\Gamma:=\frac12(A+\widetilde A)$ 
-- and the total derivative yields manifest invariance under diagonal gauge transformations. } that the target manifold has the structure of a 
${\hat p}$-suspended cotangent space
$M\cong T^\ast[\hat p]N$ 
with canonical  coordinates  
\begin{eqnarray}
 Z^i &=& (X^\alpha,P_\alpha)\ ,\qquad {\rm deg}(X^\alpha)+{\rm deg}(P_\alpha)~=~
\hat p\ ,\qquad {\rm deg}(X^\alpha)\,,
\;{\rm deg}(P_\alpha)~\in~\mathbb{N}\ .\eea
Moreover, the pre-symplectic form can be chosen to be given by\footnote{This 
choice is equivalent to using $\vartheta_{\rm alt}=\frac12 Z^i dZ^j {\cal O}_{ij}=\frac12\left(dX^\a P_\a-(-1)^{\a(\hat p+1)}dP_\a X^\a\right)$ and adding 
Gibbons--Hawking-type boundary terms of the form 
$\frac12 \sum_\xi \oint_{\partial B_\xi} X^\a P_\a$.}
\begin{eqnarray}
\vartheta &=& {\rm d}X^{\alpha}P_{\alpha}\ ,\quad {\cal O}~=~(-1)^{\a+1}
 {\rm d}X^\a {\rm d}P_\a\ ,
\quad {\cal P}~=~\tfrac{1}{2}\,\left((-1)^{\hat p \a} \partial_\a \partial^\a 
+(-1)^{\a+\hat p+1}\partial^\a \partial_\a\right)\ ,
\quad\\
{\cal O}_{ij}&=& \widetilde{\cal O}_{ij}~=~\left[\ba{cc}0&(-1)^{\a+1}\d_\a{}^\b\\ 
(-1)^{\hat p(\a+1)}\d^\a{}_\b&0\ea\right]\ ,\  {\cal P}^{ij}
~=~\left[\ba{cc}0&(-1)^{\hat p \a}\d^\a{}_\b\\ 
(-1)^{\a+\hat p+1}\d_\a{}^\b&0\ea\right]\,.\qquad\
\end{eqnarray}
The equations of motion and structure equation now read
\begin{eqnarray}
 &{\cal R}^\a ~=~ {\rm d}X^\a+{\cal Q}^\a ~\approx ~ 0\ ,\qquad 
 {\cal R}_\a ~=~ {\rm d}P_\a+{\cal Q}_\a~\approx~0\ ,
\quad&\\
&{\cal Q}^\a ~=~ (-1)^{\hat p(\a+1)+1}\partial^\a{\cal H}\ ,\qquad 
 \quad{\cal Q}_\a ~=~(-1)^{\a}\partial_\a{\cal H}\ ,
\quad&\\ 
&(-1)^\a\partial_\a {\cal H}\partial^\a{\cal H} ~\equiv~ 0\ ,
\qquad {\rm d}(\vartheta-{\cal H})~\equiv ~(-1)^{\a+1} {\cal R}^\a {\cal R}_\a\ .&
\label{dcalL}
\end{eqnarray}
The power-series expansion of ${\cal H}$ in $P_\a$ yields rank-$n$ poly-vector 
fields $\Pi{(n)}$ on $N$ of degrees $1+(1-n)\hat p$ whose mutual Schouten 
brackets vanish, \emph{viz.} 
\bea 
\{\Pi_{(n_1)},\Pi_{(n_2)}\}_{\rm S.B.}~\equiv~ 0 \quad
\mbox{for all $n_1, n_2\geqslant 0$\,.}
\eea
Using the notation $\varepsilon^i=(\e^\a,\eta_\a)$ and choosing 
$\Upsilon_\varepsilon=-\e^\a P_\a\,$, 
the gauge variation of $S^{\rm cl}_{\rm bulk}[X,P|B]$ reads
\bea 
\delta_\varepsilon {\cal L}^{\rm cl}_{\rm bulk} &=&
{\rm d}K_\varepsilon\ ,\qquad K_\varepsilon ~=~ 
(-1)^{\hat p(\a+1)}\eta_\a {\cal R}^\a
+\left((\overrightarrow P-1) \overrightarrow \e 
+\overrightarrow P\overrightarrow \eta\right){\cal H}\ ,
\eea
where we have defined
\bea \overrightarrow P~:=~ P_\a \frac{\partial}{\partial P_\a} \ ,\qquad 
\overrightarrow \e~=~ \e^\a \frac{\partial}{\partial X^\a} \ ,\qquad 
\overrightarrow \eta~=~\eta_\a \frac{\partial}{\partial P_\a} \ .\eea
Globally-defined formulations of fiber-bundle type, as discussed above, thus arise 
by using transition functions with parameters $t^{\xi'}_{\xi}=(t^\a,0)^{\xi}_{\xi'}$ 
obeying\footnote{As for (\ref{deltatdeltaZ}), it can be checked that 
$\delta_t (\delta X^\a P_\a)~=~\overrightarrow{\delta X}(\overrightarrow P-1)\overrightarrow t {\cal H}~=~0\ .$
}
\bea (\overrightarrow P-1) \overrightarrow t {\cal H}~=~0\quad 
\Leftrightarrow\quad \overrightarrow t \Pi_{(n)}~=~0\quad
\mbox{for $n\neq 1$\,,}\eea
and imposing the boundary conditions 
\bea 
\left. K_{\varepsilon}\right|_{\partial B}~\equiv~ 0\ .
\eea
The latter can be implemented by the following Dirichlet conditions: 
\bea 
\left.\eta_{\a}\right|_{\partial B}~\equiv~0\ ,\qquad 
\left.P_{\a}\right|_{\partial B}~\equiv~0\ ,
\label{Bconditions}
\eea
provided that the function 
\bea 
\Pi_{(0)}~\equiv~ {\cal H}|_{P_\a=0}~\equiv~0\ .
\label{Pi0conditions}
\eea
For these globally-defined models, the resulting integrable structures 
in the target space encompass
\begin{itemize}
 \item[(i)] a vector field $Q:=\Pi_{(1)}$ of degree $1$ that is nilpotent in the sense 
 that ${\cal L}_QQ=2\{Q,Q\}\equiv 0$, referred to as the $Q$-structure; 
 \item[(ii)] a tower of generalized Poisson structures $\Pi_{(n)}$ with 
 $n\geqslant 2$ that are compatible with $Q$ in the sense that 
 ${\cal L}_Q \Pi_{(n)}\equiv 0$;
 \item[(iii)] if in addition $\Pi_{(n)}=0$ for $n\geqslant 3$ then $\Pi_{(2)}$ is a 
 Poisson structure equipping $N$ with a Poisson bracket of intrinsic degree 
 $-\hat p + 1\,$, referred to together with $Q$ as a $QP$-structure. 
\end{itemize}

\paragraph{Transition amplitudes.}

Proceeding by assuming that $\partial B=\cup_\lambda B'_\lambda\,$, 
where $B'_\lambda$ are connected boundary components,
the space ${\cal M}$ of saddle points consists of gauge-equivalence classes of 
maps $\phi: T[1]B\rightarrow T^\ast[\hat p]N$ obeying 
${\cal R}^\a\approx0\approx {\cal R}_\a$ on $B$ 
and $P_\a|_{\partial B}\equiv 0\,$. 
Conversely, a set $\{\phi_\lambda:T[1]B'_\lambda\rightarrow N\}$ of boundary 
configurations obeying
\bea 
R^\a|_{B'_\lambda}~\approx~0\ ,\qquad R^\a~:=~{\rm d}X^\a+Q^\a\ ,
\eea
may be referred to as being (classically) compatible with $(\vartheta,{\cal H})$ 
provided there exists an extrapolating bulk manifold $B$ with  
$\partial B=\cup_\lambda B'_\lambda$ and a map 
$\phi:T[1]B\rightarrow T^\ast[\hat p]N$ obeying 
${\cal R}^\a\approx0\approx {\cal R}_\a$ on $B$ 
and $\phi|_{B'_\lambda}=\phi_\lambda\,$,
which requires generalized Poisson structures in the non-trivial case.
Semi-classically, the corresponding ``third-quantized'' transition amplitude
\bea 
{\cal A}[\phi_\lambda]~\sim~ \sum_{B} J(B) \exp\left(\frac{i}\hbar 
S^{\rm cl}_{\rm bulk}[\phi|B]\right)\ ,
\quad {\rm where}\quad \phi\vert_{B'_\lambda}=\phi_\lambda\ ,
\eea
where $J(B)$ comprises functional determinants -- combining into finite 
topological invariants once contributions from gauge-fixing sectors are included.

\paragraph{Generalized action-angles.} 

Modified amplitudes arise upon perturbing $S^{\rm cl}_{\rm bulk}$ by topological 
vertex operators which are functionals $\oint_{C} {\cal V}(X,dX)$ obeying
\bea 
\delta {\cal V}(X,dX)~=~ \delta X^\a M_{\a\b}(X,dX) R^\b
+{\rm d}(\delta X^\a {\cal P}_\a(X,dX))\ ,\label{TVO}\eea
for some matrices $M_{\a\b}$\,.
Adding such perturbations with $C\subseteq \partial B$ to $S^{\rm cl}_{\rm bulk}$ yields a modified action
\bea S^{\rm cl}_{\rm tot}[X,P;\mu_i|B;C_i]~:=~S^{\rm cl}_{\rm bulk}[X,P|B]+\sum_r \mu_r\int_{C_r} {\cal V}^r\ ,\qquad C_r~\subseteq~\partial B\ ,\eea
where $\mu_r$ are parameters. 
The total variation of the action now consists of bulk terms, which impose ${\cal R}^\a\approx0\approx {\cal R}_\a$\,, plus boundary terms that all vanish on-shell due to the boundary condition $P_\a|_{\partial B}\equiv0$ (which holds off-shell and that implies $R^\a|_{\partial B}\approx 0$).
Hence
\bea \delta \int_{C_r} {\cal V}^r~\approx~0\ ,\eea  
that is, the on-shell values of the perturbations are classical observables
\bea 
{\cal O}^r[X|C_r]~:=~\int_{C_r} {\cal J}^r(X)\ ,\qquad {\cal J}^r
~:=~{\cal V}^r(X^\a,-Q^\a)\ ,
\eea
that are defined intrinsically in the sense that if $\delta_{C_r}$ denotes a small variation of $C_r$ then 
\bea 
{\rm d}{\cal J}^r~\approx~0\qquad 
\Rightarrow\qquad \delta_{C_r} {\cal O}^r~\approx~0\ .
\eea
On general grounds, such functionals are locally-defined functions on ${\cal M}$ 
as their finiteness requires further boundary conditions on $X^\a|_{B'_\lambda}\,$.
Perturbatively, in weak-field expansions, the latter amount to taking linearized 
boundary zero-form integration constants and gauge-functions in suitable 
representations $R_\Sigma$ of the underlying Cartan gauge algebra; for related 
analyses in the case of higher-spin gravity, see \cite{Iazeolla:2011cb}.
In other words, finiteness of ${\cal O}^r$ holds in a super-selection sector given by 
a region of ${\cal M}$ labelled by a set $\{R_\Sigma\}$ of representations of the 
gauge algebra.  
One may refer to a set $\{\int_{C_r} {\cal V}^r\}$ of topological vertex operators as 
being complete if $\{{\cal O}^r\}$ is a set of (locally-defined) coordinates on (a 
super-selection sector of) ${\cal M}\,$.

Treating $\mu_r$ as generalized chemical potentials leads to the notion of a 
grand canonical ensemble with partition function 
\bea 
Z\{\mu_r;w\}~=~\left\langle \prod_r e^{\frac{i\mu_r}\hbar 
\int_{C_r}{\cal V}^r}\right\rangle_{B}\ ,
\eea
where $w$ denotes the moduli hidden in the transition functions and 
\bea 
\langle (\cdot)\rangle_B~\sim~\int {\cal D}X {\cal D}P  (\cdot) 
e^{\frac{i}\hbar S^{\rm cl}_{\rm bulk}[X,P|B]}\ ,
\label{PImeasure}
\eea
denotes a suitably regularized path integral measure (to be out-lined below). 
Micro-canonical ensembles with fixed $\int_{C_r} {\cal V}^r=q^r$ are then 
described by partition functions 
\bea 
\widetilde Z\{q^r;w\}~=~ \prod_r \int{d\mu_r\over 2\pi}  
e^{-\frac{i q^r \mu_r}\hbar}  Z\{\mu_r;w\}\ ,\eea
given by path integrals with fixed boundary conditions, \emph{viz.}
\bea 
\widetilde Z\{q^r\}~\sim~ \left\langle \prod_r \delta\left(\int_{C_r} 
{\cal V}^r-q^r\right) \right\rangle_{B}\ .
\eea
The open Poisson sigma model can be made closed by filling in the boundary 
components $B'_\lambda$ with open bulk manifolds $B_\lambda$ 
obeying $\partial B_\lambda=-B'_\lambda\,$, 
which may require additional transition functions introducing further moduli that 
we denote by $w'$, and considering the partition function
\bea 
\overline Z\{\mu_r;w,w'\}~ := ~ \left\langle \prod_r e^{\frac{i\mu_r}\hbar 
\int_{C_r} {\cal V}^r}\right\rangle_{\overline B}\ ,\qquad \overline B
~:=~B\cup \bigcup_\lambda B_\lambda\ . 
\eea
In the semi-classical limit, the filled-in bulk actions 
$S^{\rm cl}_{\rm bulk}[X,P|\overline B]$ become total derivatives 
(depending on $w'$) which may play the role of counter-terms possibly along 
the lines of the recent work in \cite{Perez:2012cf}.

\subsection{BV master action}

\paragraph{AKSZ quantization.}

The path integral measure (\ref{PImeasure}) can be defined using the BV 
field-anti-field formalism following the AKSZ approach -- see \emph{e.g.} 
\cite{Roytenberg:2006qz} for a review and references. 
To this end, the first step is to extend the classical sigma model by introducing 
layers of ghosts 
in correspondence with the classical gauge structure.
The first layer of ghosts, which have the same form degree as the gauge 
parameters,
have their own gauge symmetries, corresponding to gauge-for-gauge symmetries,
which induce a second layer of ghosts, or ghost-for-ghosts, with one unit less of 
form degree, and so on.
Proceeding this way, via a canonical procedure to be reviewed below, yields a 
minimal quantum sigma model in which all fields have non-negative ghost 
numbers and which exhibits the complete gauge structure.
As for the second step, which is the actual gauge-fixing procedure, involving the 
pairing of ghosts with suitable ghost-momenta and the introduction of Lagrange 
multipliers,
it need not be unique, as various gauge-fixing schemes may refer to different 
additional special structures in target space over and above the generalized 
Poisson structures going into the (unique) minimal model.
We shall not enter any further into these details but simply note the existence of a 
canonical (maximal) gauge-fixing scheme, that does not refer to any special 
target-space structures, with the salient features of a (classically) conserved BRST 
current and vacuum-bubble cancellation \cite{Ikeda:2012pv}.

In order to arrive at the minimal quantum model, the classical map 
$\phi: T[1]B\rightarrow M$ 
is extended into 
\bea 
\boldsymbol{\phi}~:~T[1]B\rightarrow \boldsymbol{M}\ ,
\eea
where $\boldsymbol{M}$ is a bi-graded symplectic manifold containing $M$ as a 
sub-manifold. As observed by AKSZ, the symplectic structure on $M$ induces 
dittos on $\boldsymbol{M}$ and ${\rm Maps}\left[T[1]B,\boldsymbol{M}\right]$ 
with the graded Poisson bracket of the latter being the BV bracket 
$(\cdot,\cdot)\equiv (\cdot,\cdot)_{\rm BV}\,$, the basic geometric structure 
underlying the BV field-anti-field formalism. 
Thus, in a certain space of local and ultra-local superfunctionals, based on a 
suitable extension of $\O(M)$ into $\Omega(\boldsymbol{M})\,$, 
the BV bracket is equivalent to the graded Poisson bracket on $M\,$. 
Moreover, the BV bracket-adjoint action of the integral of the pre-symplectic 
form on $\boldsymbol{M}$ over $T[1]B$ generates the exterior derivative. 
Taken together, these two lemmas imply that the classical BV master equation 
$(S,S)=0\,$, 
subject to the functional boundary condition that $S$ reduces to $S^{\rm cl}$ as 
all anti-fields are set to zero, has a simple and elegant solution given by the AKSZ 
action $\boldsymbol S\,$, which then also solves the quantum master equation, 
as we shall review next.

\paragraph{Vectorial superfields.}

Each classical coordinate $Z^i\equiv Z^{i\;\langle0\rangle}_{[p_i]}$ on $M$ is 
extended into a tower of coordinates and conjugated anti-coordinates on 
$\boldsymbol{M}$ as follows:
\begin{eqnarray}
&\left\{
{Z}^{i \;\langle g \rangle}_{[p_i-g]}\  ,\qquad 
Z^{\langle -1-g\rangle}_{i[\hat p+1-p_i+g]}~:=~
\left({Z}^{i \;\langle g \rangle}_{[p_i-g]}\right)^+ \right\} \ ,
\qquad g~=~0,\dots,p_i\ ,&
\\
&|{Z}^{i \;\langle  g \rangle}_{[p_i-g]}|~=~ 
p_i\ ,\qquad |{Z}^{\langle -1-g  \rangle}_{i[\hat p+1-p_i+g]}|~=~
\hat p-p_i\ ,&
\end{eqnarray} 
where ${O}^{\langle g \rangle}_{[p]}$ denotes a component with distinct ghost 
number $g$ and form degree $p\,$. 
The total degree and Grassmann parity (for classical theories consisting of only 
bosonic fields) are defined, respectively, by
\begin{eqnarray}
|\cdot| ~:=~ {\rm deg}(\cdot) + {\rm gh} (\cdot)\ ,\quad 
 {\rm Gr}(\cdot)~=~|\cdot|\quad{\rm mod}~2\ .
\end{eqnarray}
Given a differential form $L \in \Omega(\boldsymbol{M})$ of fixed total degree 
$|L|\,$, 
described locally on $\boldsymbol{M}$ by a function $L(Z,Z^+,dZ,dZ^+)\,$, 
with pull-back $\pi\boldsymbol{\phi}^\ast(L)\equiv \sum_{p=0}^{\hat p+1}\left[\pi\boldsymbol{\phi}^\ast( L)\right]^{\langle |L|-p\rangle}_{[p]}\in \Omega(B)$ 
and a $p$-cycle $C\subseteq B\,$, the integral 
\begin{equation}
I(L|C)~ \equiv~ \sum_\xi \int_{ B_\xi\cap C} \pi\,\boldsymbol{\phi}^*_\xi( L) ~:=~ \sum_\xi \int_{ B_\xi\cap C} \left[\pi\boldsymbol{\phi}^\ast L\right]^{\langle |L|-p\rangle}_{[p]}  \quad \mbox{\emph{i.e.}}\quad {\rm gh} (I(L|C))~=~|L|-p\ .
\label{measure}
\end{equation}
The canonical coordinates $Z^i=(§X^\a,P_\a)$ of $M$ induce supercoordinates 
$ \boldsymbol{Z}^i=(\boldsymbol{X}^\a,\boldsymbol{P}_\a)$ of 
$\boldsymbol{M}$ of fixed total degree as follows:
\begin{eqnarray}
 \boldsymbol{X}^\alpha &=& \underbrace{X^{\alpha\,\langle p_\alpha \rangle}_{[0]} +
                             X^{\alpha\,\langle p_\alpha-1 \rangle}_{[1]} + \ldots +  
                             X^{\alpha\,\langle 0 \rangle}_{[p_\alpha]}}_{\footnotesize\mbox{fields}} + 
                             \underbrace{P^{\alpha\,\langle -1 \rangle}_{[p_\alpha+1]} +
                             P^{\alpha\,\langle -2 \rangle}_{[p_\alpha+2]} + \ldots +  
                             P^{\alpha\,\langle p_\alpha - \hat p - 1 \rangle}_{[\hat p + 1]}}_{\footnotesize\mbox{anti-fields}}
\label{mathbfX}\quad,\\
\boldsymbol{P}_{\alpha} &=& \underbrace{
                               P^{\langle \hat p - p_\alpha \rangle}_{\alpha\;\,[ 0]} + 
                               P^{\langle \hat p - p_\alpha-1 \rangle}_{\alpha\;\,[1]} +\ldots +
                               P^{\langle 0 \rangle}_{\alpha\;\,[\hat p - p_\alpha]}
                    }_{\footnotesize\mbox{fields}} + 
                             \underbrace{X^{\langle -1 \rangle}_{\alpha\;\,[ \hat p - p_\alpha +1]} + 
                               X^{\langle -2 \rangle}_{\alpha\;\,[\hat p - p_\alpha +2]} +\ldots +
                               X^{\langle -p_\alpha-1 \rangle}_{\alpha\;\,[\hat p +1]}}_{\footnotesize\mbox{anti-fields}}
\label{mathbfP}\quad.\qquad
\end{eqnarray}
In these coordinates, the symplectic and pre-symplectic forms $\boldsymbol{O}$ 
and $\boldsymbol{\vartheta}$, respectively, on $\boldsymbol{M}$ read
\begin{equation}
\boldsymbol{O}~=~\left[(-1)^{\a+1} {\rm d}\boldsymbol{X}^\a 
{\rm d}\boldsymbol{P}_\a\right]^{\langle 0\rangle}_{[\hat p+2]}
~=~{\rm d}\boldsymbol{\vartheta}\ ,\qquad \boldsymbol{\vartheta}
~=~\left[ {\rm d}\boldsymbol{X}^\a 
\boldsymbol{P}_\a\right]^{\langle 0\rangle}_{[\hat p+1]}\ ,
\end{equation}
and we denote the corresponding graded Poisson bracket on $\boldsymbol{M}$ 
by 
\bea
\{\cdot,\cdot\}~\equiv~\{\cdot,\cdot\}^{\langle 0\rangle}_{[-\hat p]}\ ,
\eea
which thus has intrinsic quantum numbers 
${\rm gh}\left(\{\cdot,\cdot\}\right)=0$ and 
${\rm deg}\left(\{\cdot,\cdot\}\right)=-\hat p$\,.
The evaluation maps 
$\boldsymbol{Z}^i(p):\boldsymbol\phi \in {\rm Maps}\left[T[1]B,\boldsymbol{M}\right]\mapsto ( \boldsymbol\phi^\ast \boldsymbol{Z}^i)(p)$ 
for fixed $p\in T[1]B$ (see Appendix) define canonical coordinates on 
${\rm Maps}\left[T[1]B,\boldsymbol{M}\right]$ in which its symplectic form  
\bea 
\boldsymbol{\Omega}(\delta \boldsymbol{Z}_1,\delta\boldsymbol{Z}_2) ~=~
I\left((-1)^{\a+1} \delta\boldsymbol{X}^\a_1 \delta\boldsymbol{P}_{2\a}|B\right) 
- (1\leftrightarrow 2)\ ,\qquad {\rm gh} \left(\boldsymbol{\Omega}\right)
~=~-1\ ,
\label{BVbrack}
\eea
where $\delta\boldsymbol{Z}$ denotes a vector field on 
${\rm Maps}\left[T[1]B,\boldsymbol{M}\right]$ of total degree zero with 
component expansion
\bea 
\left.\delta\boldsymbol{Z}\right|_{\boldsymbol{\phi}}&=& 
\int_{p\in T[1]B}\sum_{k=0}^{\hat p+1}\left[\pi\left(\boldsymbol{\phi}^\ast 
\left( \delta{Z}^{i\langle p_i-k\rangle}_{[k]}\right)(p)\right)\left. 
\frac{\delta}{\delta Z^{i\langle p_i-k\rangle}_{[k]}{(p)}}
\right|_{\boldsymbol{\phi}}\right.
\nonumber\\[5pt]&& 
\left.+\pi\left(\boldsymbol{\phi}^\ast \left(\delta{Z}^{+\langle 
\hat p-p_i-k\rangle }_{i[k]}\right)(p)\right)\left. 
\frac{\delta}{\delta Z^{+\langle \hat p-p_i-k\rangle}_{i[k]}(p)}
\right|_{\boldsymbol{\phi}}\right]\ .
\eea
The corresponding graded Poisson bracket on 
${\rm Maps}\left[T[1]B,\boldsymbol{M}\right]\,$, 
referred to as the BV bracket, is denoted by 
\bea (\cdot,\cdot)~\equiv~(\cdot,\cdot)^{\langle 1\rangle}_{[0]}\ ,\eea
which thus has intrinsic quantum numbers ${\rm gh}\left((\cdot,\cdot)\right)=1$ 
and ${\rm deg}\left((\cdot,\cdot)\right)=0\,$.

\paragraph{BV bracket induced from Poisson bracket.}

As observed by AKSZ, the BV bracket $(\cdot,\cdot)$ on 
${\rm Maps}\left[T[1]B,\boldsymbol{M}\right]$ is induced from the graded Poisson 
bracket $\{\cdot,\cdot\}$ on $\Omega^{[0]}(\boldsymbol{M})$ via the formula 
\bea
  \left(\, I(F|B)\,,\,  \boldsymbol{\phi}^*(F')\,\right)~\equiv~
  \boldsymbol{\phi}^* (\{F,F'\})
\ ,
\eea
for $F,F'\in \Omega^{[0]}(\boldsymbol{M})\,$. 
It follows that the BV-adjoint action of the pre-symplectic form is related to the 
exterior derivative as follows:
\bea 
\left( \,I({\rm d}\boldsymbol{X}^{\alpha}\boldsymbol{P}_{\alpha}|B)\,,\,  
\boldsymbol{\phi}^*(L)\,\right)~\equiv~{\rm d}\,
\boldsymbol{\phi}^*(L)~\equiv~\boldsymbol{\phi}^*({\rm d}\,L)\ ,
\eea
for $L\in \Omega(\boldsymbol{M})$\,. 
We note that $\boldsymbol{\phi}^*(L)$ is an ultra-local functional, 
\emph{i.e.} a function on ${\rm Maps}\left[T[1]B,\boldsymbol{M}\right]\,$, 
\emph{idem} $F,F'\in \Omega^{[0]}(\boldsymbol{M})$\,, 
and that since ${\rm deg}({\rm d})=1$ and ${\rm gh}({\rm d})=0$ one has 
$\left( I({\rm d}\boldsymbol{X}^{\alpha}\boldsymbol{P}_{\alpha}|B),  \boldsymbol{\phi}^*(L^{\langle g\rangle}_{[p]})\right) \equiv  \boldsymbol{\phi}^*({\rm d}L^{\langle g+1\rangle}_{[p-1]})\,$. 

\paragraph{Superfunctionals} are functionals built from ultra-local 
superfunctionals $\boldsymbol{\phi}^\ast(\boldsymbol{G})$ 
where $\boldsymbol G\in \Omega(\boldsymbol{M})$ have local representatives of 
the form 
$\boldsymbol G=G(\boldsymbol{Z}^i,d \boldsymbol{Z}^i)$ 
where $G\in \Omega(M)\,$. 
In particular, if $\boldsymbol F,\boldsymbol F'$ are superfunctions 
it follows that 
\bea 
\{\boldsymbol F,\boldsymbol F'\}~=~\left.\left(\{F,F'\}_{[-\hat p]}(Z^i)\right)
\right|_{Z^i\rightarrow \boldsymbol{Z}^i}\ ,
\eea
where $\{F,F'\}_{[-\hat p]}$ denotes the Poisson bracket evaluated in the 
classical target space $M\,$. 

\paragraph{The AKSZ action}
\hspace{-.3cm} is given by the superfunctional
\begin{equation}
 \boldsymbol{S}_{\rm bulk}[\boldsymbol{\phi}|B] 
 \;:=\; I\left(\boldsymbol{L}|B\right)~=~\sum_\xi \int_{B_\xi} \pi 
 \boldsymbol{\phi}^\ast_\xi\left( \boldsymbol{L} \right)
 \ ,\qquad \boldsymbol{L}~:=~
 {\rm d}\boldsymbol{X}^{\alpha}\boldsymbol{P}_{\alpha}
 - {\cal H}(\boldsymbol{X},\boldsymbol{P})\ ,\label{minimalS}
\end{equation}
with ${\cal H}$ being a solution to the classical structure equation (\ref{structure}) 
obeying $\left.{\cal H}\right|_{P_{\a}=0}=0\,$. 
Defining
\begin{eqnarray}
{\rm s}(\cdot)~:=~(\boldsymbol{S}_{\rm bulk},(\cdot))\ ,
\end{eqnarray} 
one has
\begin{eqnarray}
{\rm s}\boldsymbol{Z}^i ~ =~{\boldsymbol{ R}}^i\ ,
\end{eqnarray}
where the generalized supercurvatures 
\begin{eqnarray}
\boldsymbol{R}^i~:=~ {\rm d}\boldsymbol{Z}^i
 + \boldsymbol{Q}^i\ ,\qquad 
 \boldsymbol{Q}^i  ~:=~{\cal Q}^i(\boldsymbol{Z}^j)~=~ 
 (-1)^{\hat p+1}{\cal P}^{ij}\partial_j {\cal H}(\boldsymbol{Z}^i)\ ,
\end{eqnarray}
with $\boldsymbol{Q}^i$ being the superfield extension of the classical 
Hamiltonian $Q$-structure in (\ref{Qstruct}). 
The locally-defined field configurations form equivalence classes modulo gauge 
transformations 
\bea 
\delta_{\boldsymbol{\varepsilon}} \boldsymbol{Z}^i~:=~{\rm d}
{\boldsymbol{\varepsilon}}^i- {\boldsymbol{\varepsilon}}^j \partial_j 
\boldsymbol{Q}^i\ ,
\eea
where the parameters have total degree 
$|\boldsymbol{\varepsilon}^i|=|\boldsymbol{Z}^i|-1$ and expansions into 
components with fixed ghost numbers and form degrees given by the suspension 
of Eqs. (\ref{mathbfX}) and (\ref{mathbfP}) with one unit of form degree, and zero 
units of ghost number. As in the classical case, it follows from 
\bea 
\delta_{\boldsymbol{\varepsilon}} \boldsymbol{S}_{\rm bulk}~\equiv~ 
\sum_{\xi}\oint_{\partial B_{\xi}} \boldsymbol{K}_{\boldsymbol{\varepsilon}}\ ,
\qquad  \boldsymbol{K}_{\boldsymbol{\varepsilon}} ~=~  
(-1)^{\hat p(\a+1)}\boldsymbol{\eta}_\a \boldsymbol{R}^\a+
\left((\overrightarrow {\boldsymbol{P}}-1) \overrightarrow {\boldsymbol{\e}} 
+\overrightarrow{\boldsymbol{P}}\overrightarrow{\boldsymbol{\eta}}\right)
{\cal H}\ ,
\eea
that the AKSZ action can be defined globally using fiber-bundle type geometries 
in which 
\begin{itemize}
\item[(i)]
the local representatives $\boldsymbol{Z}^i_\xi$ are glued together using 
transition functions with parameters 
$\boldsymbol{t}^{i,\xi}_{\xi'}=(\boldsymbol{t}^\a,0)^{\xi}_{\xi'}$ obeying 
\bea 
(\overrightarrow {\boldsymbol{P}}-1) 
\overrightarrow {\boldsymbol{t}}{\cal H}~\equiv~ 0 \quad 
\mbox{ \emph{i.e.} 
\quad $\overrightarrow{\boldsymbol t} \Pi_{(n)}\equiv 0\;$ for $\;n\neq 1$\,,}
\eea
and
\item[(ii)]
the following Dirichlet conditions are imposed:
\bea
 \left.\boldsymbol\eta_\a\right|_{\partial B}~=~0\ ,\qquad \left.
 \boldsymbol P_{\a}\right|_{\partial B}~=~0\ .
 \label{BC2}
 \eea
\end{itemize}
The AKSZ relation between the BV bracket and the Poisson bracket given above 
implies that 
\bea 
\left(\boldsymbol{S}_{\rm bulk},\boldsymbol{S}_{\rm bulk} \right) ~=~
(-1)^{\hat p}\sum_\xi \oint_{\partial B_\xi} \pi \boldsymbol{\phi}^\ast_\xi 
\left(\boldsymbol{R}^{\alpha}\boldsymbol{P}_{\alpha}-2\boldsymbol{L}\right)
~=~0\ ,
\label{SkommaS}\eea
where the latter equality follows from (\ref{BC2}) and the facts that 
$\delta_{\boldsymbol t}\boldsymbol{ L}\equiv \boldsymbol K_{\boldsymbol t}\equiv0$ 
and that 
\bea 
\delta_{\boldsymbol t} 
{\boldsymbol P}_\a~=~-(-1)^{\a}\overrightarrow{\boldsymbol t} 
\partial_\a {\cal H}\ ,\qquad
\delta_{\boldsymbol t} \boldsymbol R^\a~=~
(-1)^{\hat p(\a+1)}\overrightarrow{\boldsymbol R}_X 
\overrightarrow{\boldsymbol  t} \partial^\a {\cal H}\ ,
\eea
where we have defined 
$\overrightarrow{\boldsymbol R}_X :={\boldsymbol R}^\a \partial_\a\,$, 
which implies  
\bea 
\delta_{\boldsymbol t} (\boldsymbol R^\a 
{\boldsymbol P}_\a)~\equiv~\overrightarrow{\boldsymbol R}_X 
\overrightarrow{\boldsymbol t} (\overrightarrow{\boldsymbol P}-1){\cal H}~ 
\equiv~0\ .
\eea
In other words, the AKSZ action $\boldsymbol{S}_{\rm bulk}$ solves the classical 
BV master equation  
\begin{equation}
\left(\boldsymbol{S}_{\rm bulk},\boldsymbol{S}_{\rm bulk} \right) ~=~ 0 \quad 
\Leftrightarrow\quad {\rm s}^2~=~0\ ,
\end{equation}
subject to the functional boundary condition 
\bea 
\left.\boldsymbol{S}_{\rm bulk}[\boldsymbol{\phi}|B]
\right|_{\boldsymbol{\phi}=\phi}~=~S^{\rm cl}_{\rm bulk}[\phi|B]\ .
\eea

\paragraph{Quantum master equation.}

A remarkable property of the AKSZ formalism is that any local super-functional 
$\boldsymbol{L}$ obeys
\bea 
\Delta \boldsymbol{L}~=~0\ ,\label{DeltaL}
\eea
where $\Delta$ is the BV-Laplacian. It follows that $\boldsymbol{S}_{\rm bulk}$ 
obeys both classical and quantum master equations 
(see e.g. \cite{Ikeda:2012pv} and refs. therein), \emph{viz.}
\bea (\boldsymbol{S}_{\rm bulk},\boldsymbol{S}_{\rm bulk})~=~0~=~
 \Delta \boldsymbol{S}_{\rm bulk}\ .
 \eea
Hence $D\boldsymbol Z \exp(\frac{i}\hbar \boldsymbol{S}_{\rm bulk})$ defines a 
BRST-invariant path-integral measure (on suitable Lagrangian submanifolds): 
The classical BRST transformation $\delta_{BRST}{\cal O}:=\epsilon\, s({\cal O})\,$, 
with constant fermionic parameter $\e$ with ${\rm gh}(\e)=-1\,$, 
leaves both gauge-fixed action and $D\boldsymbol Z$ invariant; the former 
invariance requires the classical master equation while the latter invariance 
requires\footnote{More generally, it follows from (\ref{DeltaL}) that any canonical 
transformation, \emph{viz. }
$\delta_{\boldsymbol E}{\cal O}:=({\boldsymbol E},{\cal O})$ 
with ${\rm gh}(\boldsymbol E)=-1\,$, 
leaves $D\boldsymbol Z$ invariant. } 
$\Delta \boldsymbol{S}_{\rm bulk}=0\,$.
The quantization thus deforms the classical differential algebra with differential 
d and $Q$-structure $\boldsymbol Q\,$, 
which one may view as a first-quantized algebra, 
into a second-quantized operator algebra with BRST current 
$\boldsymbol j_{\rm BRST}$ (which is conserved on shell barring anomalies)
and differential ${\rm ad}_{\boldsymbol q}$ where 
$\boldsymbol q:=\oint \boldsymbol j_{\rm BRST}\,$. 
Thus, acting on second-quantized ultra-local superfunctionals $\boldsymbol F\,$, 
one has ${\rm ad}_{\boldsymbol q} \boldsymbol F={\rm d}\boldsymbol F+ \rho(\boldsymbol Q)\boldsymbol F$ 
where $\rho(\boldsymbol Q)$ denotes the realization of $\boldsymbol Q$ 
in the second-quantized algebra, that, on general grounds, carries the structure 
of a graded homotopy-associative differential algebra. 

\paragraph{Deformed master action.}

The BRST cohomology at ghost number zero consists of on-shell 
gauge-invariant observables. \cite{Barnich:2000zw} 
\footnote{At the level of locally-defined densities, one has that 
$H^{\langle g\rangle}(s)\cong H^{\langle g\rangle}(\gamma,H(\delta))$ 
where $\gamma$ generates the classical gauge symmetries and $\delta$ is the 
Koszul-Tate differential implementing the equations of motion 
\cite{Henneaux:1992ig}. 
The construction of $H^{\langle 0\rangle}(s)$ then passes via 
globally-defined formulations distinguishing between manifest off-shell 
gauge symmetries and non-manifest Cartan gauge symmetries on shell 
\cite{Boulanger:2011dd,Sezgin:2011hq}.}
Although the latter can be extended into off-shell functionals in various ways, the 
super-field  framework leads to a unique extension: Given a set of classical 
observables, $\{{\cal O}^r\}$ say, with super-field extensions 
${\boldsymbol{O}}^r={\cal O}^r[\boldsymbol Z]$ obeying 
${\rm s}\boldsymbol{O}^r=0\,$, one seeks further off-shell extensions 
\bea 
\widehat {\boldsymbol O}^r~:=~\boldsymbol{O}^r + \int_{C_r} 
\boldsymbol R^r \boldsymbol L_r \ ,\qquad {\rm s} \boldsymbol L_r ~=~0\ ,
\qquad 
(\widehat {\boldsymbol O}^r,\widehat {\boldsymbol O}^r)~=~0\ ,
\eea
\emph{i.e.} $\widehat {\boldsymbol O}^r=\boldsymbol{O}^r + {\rm s} 
\left( \int_{C_r} \boldsymbol Z^r \boldsymbol L_r\right)\,$. 
The total master action 
\bea
\boldsymbol{S}_{\rm tot}~:=~ \boldsymbol{S}_{\rm bulk}
 +\sum_r \mu_r  \widehat{\boldsymbol O}^r
\eea
then obeys the classical master equation. 
As for boundary conditions \cite{Hofman:2002rv}, the undeformed ones 
(\ref{BC2}) (imposed off shell in order to have a globally-defined bulk 
action) are compatible with those following the variational principle 
provided that the off-shell extensions are super-field extensions 
$\boldsymbol V^r ={\cal V}^r(\boldsymbol{X},{\rm d}\boldsymbol{X})$ of 
topological vertex operators as defined in (\ref{TVO}), \emph{i.e.} 
\bea \boldsymbol{S}_{\rm tot}[\boldsymbol{X},\boldsymbol{P};
\mu_i|B;C_i]~:=~\boldsymbol{S}_{\rm bulk}[\boldsymbol{X},
\boldsymbol{P}|B]+\sum_r \mu_r \int_{C_r}  {\cal V}^r(\boldsymbol{X},
{\rm d}\boldsymbol{X})\ ,
\eea
where $C_r\subseteq\partial B\,.$

\section{Vasiliev's theory: a graded-associative non-commutative 
case}\label{sec:noncommut}

In this section we begin by reviewing selected features of the action principle for 
Vasiliev's theory in four dimensions given in \cite{Boulanger:2011dd}.
We then construct a minimal classical BV master action using a natural 
generalization of the AZSZ formalism to the case of graded associative differential 
algebras. 
In addition, we shall refine the analysis of \cite{Boulanger:2011dd} concerning 
compatibility conditions for globally-defined formulations of fiber-bundle type at 
the level of classical action as well as classical BV master actions.

Before turning to the details, we wish to emphasize that while the BV anti-bracket 
generalizes straightforwardly to the non-commutative case, 
the corresponding generalization of the BV Laplacian requires the introduction of 
distributive two-point functions (delta functions) on non-commutative manifold, 
that we defer to a future work together with the analysis of the quantum BV master 
equation.
It is natural, however, to expect that the classical BV master action principle 
presented here also solves the quantum master equation.

\subsection{Classical theory}
 
\paragraph{Correspondence space.}
Vasiliev's formulation of higher-spin gravities is in terms of associative differential 
algebras on non-commutative \emph{correspondence spaces} 
$\mathfrak{C}\cong T^\ast \mathfrak{M}$ introducing the following basic 
notions:
\begin{itemize} 
\item the differential algebra ${\Omega}(\mathfrak{C})$ with differential d and 
compatible graded-associative product $\star$\,, \emph{i.e.} if 
$f,g\in \Omega(\mathfrak{C})$ then 
${\rm d}(f\star g)=({\rm d}f)\star g+(-1)^{{\rm deg}(f)}f\star({\rm d}g)\,$. 
These two operations are assumed to be real in the sense that there exists an 
anti-linear anti-automorphism $\dagger$ obeying 
$(f\star g)^\dagger =(g^\dagger)\star (f^\dagger)$ and 
$({\rm d}f)^\dagger={\rm d}(f^\dagger)$ for all 
$f,g\in \Omega(\mathfrak{C})\,$;
\item a graded cyclic trace operation 
Tr$:\Omega(\mathfrak{C})\rightarrow \mathbb{C}$ obeying 
${\rm Tr}[{\rm d}(\cdot)]\equiv0$ (modulo boundary terms), given essentially by 
the integral over $\mathfrak C$, that defines a non-degenerate bi-linear form 
compatible with d and $\star$;
\item a subalgebra consisting of d-closed central elements $J^r$, \emph{i.e.} 
${\rm d}J^r=0$ and $J^r\star f=f\star J^r$ for all $f\in\Omega(\mathfrak{C})$\,;
\end{itemize}
In the case of four-dimensional bosonic higher-spin gravities, including the 
minimal bosonic models, the differential forms take their values in the algebra 
$\mathbb Z_2 \times \mathbb Z_2$ generated by two outer Kleinians $(k,\bar k)$ 
obeying
\bea 
k\star k~=~1\ ,\qquad [k,{\rm d}]_\star~=~0\ ,\qquad k^\dagger~=~\bar k\ .
\eea
The subalgebra of d-closed central elements is generated by various projections of 
the symplectic form on $\mathfrak{C}$ together with the elements  
\begin{eqnarray}
( J^{I}_{[2]})_{I=1,2}&=&-\tfrac{i}{4}\,(1 \,,\, k \kappa )\,\star P_+\star d^2z  
\;, \nonumber \\
( J^{\bar I}_{[2]})_{\bar I=\bar 1,\bar 2} &=& 
-\tfrac{i}{4}\,( 1\,,\, \bar k {\bar\kappa} )\, \star P_+ \star d^2\bar z  \;,
\nonumber \\
P_+ &=& \tfrac{1}{2}\,(1+k\bar{k})\,,\quad \;
\label{central1}
\end{eqnarray}
where the two inner Kleinians
\bea 
\kappa~:=~(2\pi)^2 \delta^2(y)\star \delta^2(z)\ ,\qquad 
\bar\kappa~:=~(\kappa)^\dagger~=~(2\pi)^2 \delta^2(\bar y)\star 
\delta^2(\bar z)\ ,
\eea
using Weyl-ordered symbols, and $(y^\a,\bar{y}^{\ad}; z^\a,\bar{z}^{\ad})$ 
(with $\a,\ad=1,2$) are local coordinates on the doubled twistor space 
$\mathfrak{Z}_\xi\times \mathfrak{Y}_\xi\subset \mathfrak{C}$ 
obeying 
\bea 
\{k,y^\a\}_\star~=~\{k,z^\a\}_\star~=~0~=~[k,\bar{y}^{\ad}]_\star~=~
[k,\bar{z}^{\ad}]_\star\ ,\eea
\bea [y^\a,y^\b ]_\star ~=~ 2{\rm i}\e^{\a\b}\ ,\quad 
[z^\a,z^\b]_\star~ =~ -2{\rm i}\e^{\a\b}\ ,\eea
and the reality conditions 
\bea (y^\a)^\dagger~=~\bar{y}^{\ad}\ ,\qquad (z^\a)^\dagger~=~\bar{z}^{\ad}\ .\eea
The inner Kleinian obey $\kappa\star\kappa=1$ and
\bea \kappa\star f\star \kappa ~=~(-1)^{{\rm deg}_{{\mathfrak{y}}\times {\mathfrak{z}}}(f)} \pi(f)\ ,\qquad \pi(f)~:=~k\star f\star k\ ,\eea
where ${\rm deg}_{{\mathfrak{y}}\times {\mathfrak{z}}}(f)$ denotes the holomorphic form degree of $f$\,, \emph{idem} $\bar\kappa$ and $\bar\pi(f):=\bar k\star f\star \bar k$\,. 
The full correspondence space is thus of the form
\bea \mathfrak{C}~=~\bigcup_\xi {\mathfrak C}_\xi\ ,\qquad {\mathfrak C}_\xi~=~T^\ast\mathfrak M_{\xi}\times {\mathfrak{Z}}_\xi\times {\mathfrak{Y}}_\xi\ ,\eea
where $T^\ast\mathfrak{M}_\xi$ has real canonical coordinates $(X^M,P_M)$ obeying
\bea \left[X^M,P_M\right]_\star~=~i\delta^M_N\ .\eea
Requiring $(X^M,P_M;y^\a,\bar{y}^{\ad}; z^\a,\bar{z}^{\ad})$ to commute with the line-elements $(dX^M,dP_M;dy^\a,d\bar y^{\ad};dz^\a,d\bar z^{\ad})$ it follows that the latter generate a graded commutative algebra.

\paragraph{Chiral trace operation.}

The basic chiral trace operation is defined by 
\begin{equation}
 {\rm Tr}[f]~:=~\sum_{\xi} \int_{T^\ast\mathfrak M_\xi\times {\mathfrak{Y}}\times {\mathfrak{Z}}}f|_{k= 0 = \bar k}\ ,
\end{equation}
where the integral projects onto the top form degree; the integrand should be understood as the symbol of $f$ in a suitable order\footnote{If the trace is well-defined, it does not depend on the choice of order.}; and the twistor variables are integrated along independent real contours. 
This trace operation is graded cyclic, \emph{i.e.}
\bea 
{\rm Tr}[f\star g]~=~(-1)^{{\rm deg}(f){\rm deg}(g)}{\rm Tr}[g\star f]\ .
\eea
Various other graded-cyclic trace operations can be obtained by projecting Tr. 
Inserting 
\bea 
P_\pm~=~\frac12 (1\pm k\bar k)\ ,
\eea
yields a trace that is graded cyclic and non-degenerate on the bosonic subalgebras
\bea {\cal A}_\pm~:=~\left\{~f~\in~ \Omega(\mathfrak{C}) \times \mathbb Z_2 \times \mathbb Z_2~:~ f~=~\pi\bar\pi(f)~=~P_+\star f~\right\}\ .\eea
Inserting $\Omega_{{\mathfrak Y}}:= \frac{d^2y \,d^2\yb} {(2\pi)^2}$ yields a trace that is non-degenerate on $\Omega(T^\ast\mathfrak{M}\times {\mathfrak Z})\otimes \Omega^{[0]}({\mathfrak Y})$\,. Inserting also $\Pi_{\mathfrak{M}}:=\frac{1}{n!} \epsilon^{M_1\cdots M_n}
dP_{M_1}\cdots dP_{M_n} \delta^n(P_N)$, defined using Weyl order, we obtain a reduced trace operation that remains non-degenerate on $\Omega(\mathfrak{M}\times {\mathfrak Z})\otimes \Omega^{[0]}({\mathfrak Y})$, \emph{viz.}
\bea 
\check{\rm Tr}[f]~:=~{\rm Tr}[\Pi_{\mathfrak{M}}\star \Omega_{{\mathfrak Y}}\star f]~\equiv~\sum_{\xi}\int_{\mathfrak{M}_\xi} {\rm Tr}'[f]\ ,
\eea
where the twistor-space trace operation 
\bea 
{\rm Tr}'[f]~:=~\int_{{\mathfrak Y}\times {\mathfrak Z}}
\left[\Omega_{{\mathfrak Y}}\star f|_{k=\bar k=0\,;\, dP_M=0\,;\, 
P_M=0}\right]\ .
\eea
The reduced trace remains graded cyclic, \emph{i.e.}
\bea 
\check{\rm Tr}[f\star g]~=~(-1)^{{\rm deg}(f){\rm deg}(g)}\check{\rm Tr}
[g\star f]\ .
\eea

In order to make contact with the previous section, one thus treats 
\bea 
\mathfrak{M}\times {\mathfrak Z}~\equiv~B\ ,
\eea
as the bulk manifold, hence
\bea
 \hat p~=~{\rm dim}(\mathfrak{M})+3\ ,
\eea
and ${\mathfrak Y}$ as a fiber manifold, \emph{i.e.} all quantities are expanded in 
sets $\{T_{\l}(y^\a,\bar y ^{\ad})\}$ of functions on 
${\mathfrak Y}$ treated as types forming a basis for an associative $\star$-
product algebra with coefficients in $\Omega(B)$ that remains closed under 
$\star$-product composition with $\kappa$ and $\bar\kappa$\,; for a concrete 
example of this separation of variables, see \cite{Iazeolla:2011cb}. The 
choice of types is adapted to the boundary conditions on $B$ and  may hence 
manifest various symmetry algebras, such as generalized Lorentz algebras or 
compact algebras, leading to the notion of (inverse) harmonic expansions 
\cite{Iazeolla:2008ix,Boulanger:2008kw}.  
In what follows, for the purpose of setting up the AKSZ formalism, it 
suffices, however, to treat the ${\mathfrak Y}$-dependence formally.

\paragraph{Classical action: odd-dimensional bulk.}

If ${\rm dim}(\mathfrak M)=2n+1\,$ with $n\geqslant 0$\,, that is 
$\hat p= 2n+4$\,,   
a duality extension of Vasiliev's equations of motion for four-dimensional 
higher-spin gravities, which is locally equivalent to Vasiliev's original 
equations, follows from the variational principle based on the generalized 
Hamiltonian bulk action
\begin{eqnarray}
S^{\rm cl}_{\rm bulk}[\{ A, B, U, V\}_\xi]
&=& \sum_\xi \int_{\mathfrak M_\xi}  {\rm Tr}'\left[
 U \star D B+ V\star \left( F + {\cal G}(B,U;  J^I,  J^{\bar I}, 
J^{I\bar I})\right)\right]_{\xi}\ ,\qquad
\label{SclHS} 
\end{eqnarray}
where the locally-defined master fields have decompositions under total form degree into 
\begin{equation}
    A  ~=~   A_{[1]}+   A_{[3]}+  \cdots +   A_{[2n+3]}\ ,\qquad 
    B  ~=~   B_{[0]}+   B_{[2]}+ \cdots +   B_{[2n+2]}\ ,\ee\be
    U  ~=~   U_{[2]}+   U_{[4]}+ \cdots +   U_{[2n+4]}\ ,\qquad  
    V  ~=~    V_{[1]}+   V_{[3]}+ \cdots +   V_{[2n+3]}\ .
\end{equation}
The interaction freedom in ${\cal G}$ needs to be constrained in order for the action to be  gauge 
invariant. Making the ansatz\footnote{The coupling
$\tilde f:=\partial_U \widetilde{\cal F}_0\vert_{U=0}$ determines whether the 
target space is a symplectic manifold ($\tilde f\neq 0$) or a proper Poisson manifold ($\tilde f=0$).
In the symplectic case the $U$ and $V$ variables can be integrated out after which the action becomes a boundary term while this is no longer possible in the proper Poisson case.
} 
\be 
{\cal G}~=~{\cal F}(B;  J^I,  J^{\bar I}, J^{I\bar I})+
\widetilde{\cal F}(U;  J^I,  J^{\bar I}, J^{I\bar I})\ ,
\ee
\bea
 {\cal F}&=& {\cal F}_0(B) + {\cal F}_I( B)\star  J^I_{[2]} + 
              {\cal F}_{\bar I}( B)\star J^{\bar{I}}_{[2]} 
       +  {\cal F}_{I\bar I} ( B)\star  J_{[4]}^{I\bar{I}}\ ,\\[5pt]
\widetilde{\cal F}&=& \widetilde{\cal F}_0(U)+\widetilde{\cal F}_I( U)
\star  J^I_{[2]} + 
          \widetilde{\cal F}_{\bar I}( U)\star J^{\bar{I}}_{[2]} 
       +  \widetilde{\cal F}_{I\bar I} ( U)\star  J_{[4]}^{I\bar{I}}\ ,
\eea
the following two cases yields integrable equations of motion: 
\bea \mbox{bilinear $Q$-structure}&:&{\cal F}~=~  B\star  J\ ,\qquad 
J~=~J_{[2]}+ J_{[4]}\ ,\\[5pt]
\mbox{bilinear $P$-structure}&:& \widetilde{\cal F}~=~  U\star  J'\ ,\qquad  
J'~=~J'_{[2]}+ J'_{[4]}\ ,
\eea
where the central elements are defined via 
\begin{equation}
  B \star  J_{[2]}~\equiv~ {\cal F}_I \star J^I_{[2]}
  + {\cal F}_{\bar I}\star  J^{\bar I}_{[2]}  \quad , \qquad 
   B \star J_{[4]} 
   ~\equiv~  {\cal F}_{I\bar I}\star J^{I\bar I}_{[4]}\ ,  
\end{equation}
\begin{equation}
 J_{[2]}~=~ -\tfrac{i}{4} \, \left[ {\rm d}z^2 ( b_1+b_2 \,k\,\kappa) +
                  {\rm d}\bar z^2 (b_{\bar 1}+ b_{\bar 2}\, \bar k \, 
                   {\bar\kappa})\ \right]\star \, P_+\;,
\label{J2} 
\end{equation}
\begin{equation}
 J_{[4]}~=~ -\tfrac{i}{4}\, {\rm d}z^2 {\rm d}\zb^2 
         \left[ c_{1\bar 1} + 
            c_{2 \bar 1 }\, k\,\kappa + c_{1 \bar 2}\, \bar k\,
                 {\bar \kappa} + c_{2\bar 2}\,  \kappa 
              \,{\bar \kappa}\right] \star \, P_+ \quad.
\label{J4} 
\end{equation}
Indeed, letting $Z^{ i}=( A, B, U, V)$\,, the general variation of the action reads
\begin{equation}
\delta S^{\rm cl}_{\rm bulk}\ =\ \sum_\xi \int_{\mathfrak M_\xi}  {\rm Tr}'\left[ 
\delta  Z^{ i} \star {\cal R}^{ j} \tilde{\cal O}_{ i j}
\right]+ \sum_{\xi} \int_{\partial \mathfrak M_\xi}  {\rm Tr}' \left[  U\star\delta  B
      - V\star\delta  A \right]\ , \label{totalvar}
\end{equation}
where ${\cal O}_{ i j}$ is a constant non-degenerate matrix defining the 
symplectic form of degree $\hat p+2$ on the target space and the generalized curvatures 
\be
{\cal R}^{ A} ~=~   F + {\cal F}+ \widetilde{\cal F}\ ,\qquad 
{\cal R}^{ B} ~=~   D  B + ( V\partial_{ U})\star \widetilde{\cal F}\ ,\ee
\be {\cal R}^{ U} ~=~   D  U -( V\partial_{ B})\star {\cal F}\ ,\qquad 
{\cal R}^{ V} ~=~  D  V + [ B ,  U]_\star \ ,\ee
and the bulk equations of motion ${\cal R}^i\approx 0$ are Cartan integrable for 
the above choices of ${\cal F}$ and $\widetilde{\cal F}\,$. 
As shown in \cite{Boulanger:2011dd}, the on-shell Cartan gauge 
transformations
\bea 
\delta_{ \e,\eta}  A&=& D \e^{\, A}-(\e^{\, B}\partial_{ B})\star {\cal F}
-(\eta^{\, U}\partial_{ U})\star\widetilde{\cal F}\ ,\\[5pt]
\delta_{ \e,\eta}  B&=& D \e^{\, B}-[ \e^{\, A}, B]_\star-( \eta^{\, V}\partial_{ U})
\star \widetilde{\cal F}-( \eta^{\, U}\partial_{ U})\star( 
V\partial_{ U})\star\widetilde{\cal F}\ ,
\\[5pt]
 \delta_{ \e,\eta}  U&=& D \eta^{\, U}-[ \e^{\, A}, U]_\star+( \eta^{\, V}
\partial_{ B})\star 
{\cal F}+( \e^{\, B}\partial_{ B})\star( V\partial_{ B})\star{\cal F}\ ,\\[5pt]
\delta_{ \e,\eta}  V&=& D \eta^{\, V}-[ \e^{\, A}, V]_\star-[ \e^{\, B}, U]_\star 
+[ \eta^{\, U}, B]_\star\ ,
\eea
remain symmetries off shell modulo boundary terms, \emph{viz.}
\begin{eqnarray}
\delta_{\epsilon, \eta}S^{\rm cl}_{\rm bulk}[A,B,U,V] ~=~
\sum_\xi \int_{\partial \mathfrak{M_\xi}} K_{\eta}\ ,
\end{eqnarray}
where 
\begin{eqnarray} 
K_{\eta}~=~{\rm Tr}'\left[ 
\eta^U \star DB + \eta^V \star (F + {\cal F} + (1-U\partial_U)\star
\widetilde{\cal F})\right]\ .  
\end{eqnarray}
As found in \cite{Boulanger:2011dd}, the closure formula for Cartan gauge 
transformations generalized straightforwardly from the commutative to the 
non-commutative case, \emph{i.e.}
\bea [\delta_{\vare},
\delta_{\vare}]Z^i~=~\delta_{\vare_{12}}Z^i-\overrightarrow{\cal R}\star 
\vare^i_{12}\ ,
\eea
with composite parameters
\bea 
\vare^i_{12}~=~- \overrightarrow{\vare}_1 \star 
\overrightarrow{\vare}_2 \star {\cal Q}^i\ ,
\eea
which can be used to construct globally-defined bulk actions within the context of fiber bundles. Thus, the contributions to $\delta_{\epsilon, \eta}S^{\rm cl}_{\rm bulk}$ from the chart boundaries in the interior of $\mathfrak{M}$ can be made to cancel by gluing together the locally-defined field configurations and broken $\eta$-gauge parameters using gauge transitions $\hat\delta_t Z^i=\delta_t Z^i$ and
\begin{eqnarray}
\hat\delta_t \eta^U &=& -[t^A,\eta^U] - 
(t^B\partial_B)\star(\eta^V\partial_B)
{\cal F}\quad ,
\\
\hat\delta_t \eta^V &=& -[t^A,\eta^V] + \{ \eta^U,t^B\} \quad,
\end{eqnarray}
\emph{i.e.}
\bea \hat\delta_{t} K_{\eta}~=~0\ ,\eea
where, moreover, the compatibility conditions on $\{t^A,t^B\}$ read as follows:
\bea \overrightarrow{\cal R}\star [\overrightarrow{t},\overrightarrow{\e}]_\star \star {\cal Q}^i~=~0\qquad \mbox{for all $i$, $\overrightarrow{\cal R}$ and $\overrightarrow{\e}$\,.}\label{compcondVas}\eea
The conditions on $t^A$ hold for all ${\cal F}$ while those for $t^B$ hold only if ${\cal F}$ is at most bi-linear. Thus, if ${\cal F}$ is at least tri-linear then $t^B$-transitions must be discarded. 


\paragraph{Classical action: even-dimensional bulk.}
If ${\rm dim}(\mathfrak M)=2n$ with $n\geqslant 0$\,, that is $\hat p= 2n-1$, the duality-extended equations of motion follow from the variational principle based on the generalized Hamiltonian bulk action
\be S^{\rm cl}_{\rm bulk}[ A, B; S, T]~=~  \sum_\xi \int_{\mathfrak M_\xi}  {\rm Tr}'\left[  
S \star  D  B
+ T\star ( F+{\cal F}+\widetilde {\cal F}( S;  J^I, J^{\bar I},J^{ I\bar J}))\right]_{\xi}\ ,
\label{Scleven}
\ee
where the interaction function obeys
\be 
\widetilde{\cal F}(-S)~=~\widetilde{\cal F}(S)\ ,\qquad \widetilde{\cal F}|_{S=0}~=~0\ ,
\ee
and the fields are assigned form degrees as follows:
\begin{equation}
    A  ~=~   A_{[1]}+   A_{[3]}+  \cdots +   A_{[2n-1]}\ ,\qquad 
    B  ~=~   B_{[0]}+   B_{[2]}+ \cdots +   B_{[2n-2]}\ ,\ee\be
    S  ~=~   S_{[1]}+   S_{[3]}+ \cdots +   S_{[2n-1]}\ ,\qquad  
    V  ~=~    T_{[1]}+   T_{[2]}+ \cdots +   T_{[2n-2]}\  .
\end{equation}
From the variation one obtains
\be {\cal R}^{ A}~=~ F +  {\cal F } + \widetilde{\cal F}(S)\ ,\qquad 
{\cal R}^{ B}~=~  D B -( T\partial_{ S})\star \widetilde{\cal F}( S)\ ,\ee
\be {\cal R}^{ S}~=~ D S+ (T\partial_B)\star {\cal F}\ ,\qquad 
{\cal R}^{ T}~=~ D T+[ S, B]_\star \ .\ee
and the integrability of the equation of motion $ D{\cal R}^I-({\cal R}^J\partial_J)\star Z^I~\equiv~0 $ requires
\bea
D{\cal R}^A - ({\cal R}^B \partial_B) \star {\cal F} - ({\cal R}^S \partial_S) \star \widetilde{\cal F} &=& ((T \partial_S) \star \widetilde{\cal F} \partial_B) \star {\cal F} -  ((T \partial_B) \star {\cal F} \partial_S) \star \widetilde{\cal F} ~\equiv~0\ ,  \qquad \eea \bea 
D{\cal R}^B - [ {\cal R}^A,  B] + ({\cal R}^T \partial_T) \star \widetilde{\cal F}  + ({\cal R}^S \partial_S) \star ((T \partial_S) \star \widetilde{\cal F}) &=& ((T \partial_B) \star {\cal F} \partial_S) \star (T \partial_S\widetilde{\cal F}) ~\equiv~0\ ,\qquad\quad \\
D{\cal R}^S - [ {\cal R}^A,  S] - ({\cal R}^T \partial_B) \star {\cal F}  - ({\cal R}^B \partial_B) \star ((T \partial_B) \star {\cal F}) &=& ((T \partial_S) \star \widetilde{\cal F} \partial_B) \star (T \partial_B{\cal F}) ~\equiv~0\ , \eea
whereas 
\bea 
 D{\cal R}^T - [ {\cal R}^A,  T] - [ {\cal R}^S,  B] +  \{ {\cal R}^B,  S\} \equiv 0 \ ,
\eea
as follows from the even functions $\widetilde{\cal F}$ obey 
\be 
\{  S, ( T\partial_{ S})\star \widetilde{\cal F}\}_\star~\equiv~[ T,\widetilde{\cal F}]_\star\ .
\ee
The remaining conditions are satisfied in two cases:
\be
{\cal F} = B \star f( J^I, J^{\bar I},J^{ I\bar J}) \ , \quad 
\widetilde{\cal F} = \sum_n S^{\star 2n} \star w_n( J^I, J^{\bar I},J^{ I\bar J})
\ee
or
\be
{\cal F} = \sum_n B^{\star n} \star f_n( J^I, J^{\bar I},J^{ I\bar J}) \ , \quad \widetilde{\cal F} = 0. 
\ee
where $f_n ,w_n$ are arbitrary  functions of the central terms $J^I, J^{\bar I},J^{ I\bar J}$.
This choice makes the action invariant under the gauge transformations
\bea 
\delta_{ \e,\eta}  A &=& D \e^{\, A}-(\e^{\, B}\partial_{ B})\star {\cal F}-(\eta^{\, S}\partial_{ S})\star \widetilde{\cal F}\ ,\\[5pt]
\delta_{ \e,\eta}  B&=& D \e^{\, B}-[ \e^{\, A}, B]_\star + ( \eta^{\, T}\partial_{ S})\star\widetilde{\cal F} + ( \eta^{\, S}\partial_{ S})\star( T \partial_{ S})\star\widetilde{\cal F}\ ,\\[5pt]
 \delta_{ \e,\eta}  S&=& D \eta^{\, S}-[ \e^{\, A}, S]_\star+( \eta^{\, T}\partial_{ B})\star {\cal F}+( \e^{\, B}\partial_{ B})\star( T\partial_{ B})\star{\cal F}\ ,\\[5pt]
\delta_{ \e,\eta}  T&=& D \eta^{\, T}-[ \e^{\, A}, T]_\star + \{ \e^{\, B}, S \}_\star - [  \eta^{\, S}, B]_\star \ ,
\eea
up to the boundary terms
\be
\delta_{ \epsilon,\eta} S^{\rm cl}_{\rm bulk} ~=~ \sum_\xi
\int_{\partial {\mathfrak{M}}_\xi}{\rm Tr}\left[ {\eta}^S 
\star D B 
+ \eta^T \star( F +  {\cal F} + (1-S\partial_S)\star\widetilde{\cal F})\right] \ .
\ee
The construction of a globally-defined action and the required compatibility conditions are  analogous to the case of even $\hat p$ using 
\bea 
 \hat\delta_{t}  \eta^{\, S}&=& -[ t^{\, A}, \eta^{\, S} ]_\star+( t^{\, B}\partial_{ B})\star (\eta^{\, T} \partial_B) \star {\cal F} \quad ,\\[5pt]
\hat \delta_{t}  \eta^{\, T}&=& -[ t^{\, A}, \eta^{\, T} ]_\star-[ t^{\, B}, \eta^{\, S} ]_\star \quad .
\eea
%


\subsection{AKSZ master action} 

\paragraph{The bulk action.}

In this section we give the minimal solutions $\boldsymbol{S}$
of the classical master equation corresponding to the classical 
action principles given in the previous sections. 

The classical fields $Z^i$ become coordinates $\boldsymbol{Z}^i$ of a 
supermanifold and contain all the ghosts and antifields of the 
BRST-BV spectrum similarly to what is explained below (\ref{mathbfX})
and (\ref{mathbfP}). The AKSZ master actions are obtained by taking the classical bulk actions 
(\ref{SclHS}) and (\ref{Scleven});
replacing $Z^i$ by $\boldsymbol{Z}^i$ therein; integrating as in 
(\ref{measure}) so as to select only the top form component of the 
resulting Lagrangian density; and projecting onto ghost number zero, \emph{viz.}
\bea 
\boldsymbol{S}~=~\left.S^{\rm cl}_{\rm bulk}[\boldsymbol{Z}^i]\right|^{\langle 0\rangle}~\equiv~ \sum_{\xi}\left. \int_{B_\xi} {\rm Tr}' \boldsymbol L_\xi\right|^{\langle 0\rangle} \ ,\eea 
that is
\bea \mbox{Even $\hat p$:}&& \boldsymbol{L}~=~\boldsymbol U \star \boldsymbol D \boldsymbol B +\boldsymbol V\star \left(\boldsymbol F+{\cal F}(\boldsymbol B;J^r)+\widetilde {\cal F}(\boldsymbol U;J^r)\right) \ ,\label{AKSZ1}\\[5pt]
\mbox{Odd $\hat p$:}&&\boldsymbol{L}~=~ \boldsymbol S \star \boldsymbol D \boldsymbol B +\boldsymbol T\star \left(\boldsymbol F+{\cal F}(\boldsymbol B;J^r)+\widetilde {\cal F}(\boldsymbol S;J^r)\right)\ .\label{AKSZ2}\eea
As for the BV bracket $(\cdot,\cdot)$ in non-commutative space, it is defined
analogously to (\ref{BVbrack}) and defines a derivation in the sense that for 
any local star-functional $F$ and ultra local star-functionals $A(p)$ and $B(p)\,$, 
evaluated on $p\in \mathfrak C\,$, it satisfies 
\begin{eqnarray}
(F,A(p)\star B(p)) = (F,A(p))\star B(p) + (-1)^{A(F+1)}A(p)\star (F,B(p))
\quad.  
\end{eqnarray}
Thus, similarly to the commutative case, we have the following basic BV brackets:
\begin{eqnarray}
({\boldsymbol S},\boldsymbol{Z}^i) = \boldsymbol{R}^i\quad ,
\qquad  
\boldsymbol{R}^i = {\rm{d}} \boldsymbol{Z}^i + \boldsymbol{Q}^i\ ,
\label{BVstar} 
\end{eqnarray}
and where $\boldsymbol{Q}^i={\cal Q}^i(\boldsymbol{Z}^i)\,$. 
  
It is then a direct computation to verify that the master equation
$({\boldsymbol S},{\boldsymbol S})=0$ is satisfied up to boundary terms as 
follows:
\bea \mbox{\underline{Even $\hat p\,$}:} \ ({\boldsymbol S},{\boldsymbol S})
&=& -\oint_{\partial B}{\rm{Tr}}'
\left[  
\boldsymbol{U} \star \boldsymbol{DB} + \boldsymbol{V}\star (\boldsymbol{F}+{\cal F}(\boldsymbol{B};J)) 
+ \boldsymbol{V}\star (1-\boldsymbol{U}\partial_{\boldsymbol{U}})\star \widetilde{\cal F}
(\boldsymbol{U};J)
\right]\,,\qquad
\label{masterodd}
\\[5pt]
\mbox{\underline{Odd $\hat p\,$}:} \ ({\boldsymbol S},{\boldsymbol S})&=& \oint_{\partial B}{\rm{Tr}}'
\left[  
\boldsymbol{S}\star \boldsymbol{DB} + \boldsymbol{T}\star (\boldsymbol{F}+{\cal F}(\boldsymbol{B};J)) 
+ \boldsymbol{T}\star (1-\boldsymbol{S}\partial_{\boldsymbol{S}})\star \widetilde{\cal F}
(\boldsymbol{S};J)
\right]\quad
\label{mastereven}
\eea
which one indeed identifies as the non-commutative generalization of (\ref{SkommaS}), \emph{i.e.}
\bea ({\boldsymbol S},{\boldsymbol S})~=~(-1)^{\hat p}\sum_\xi \oint_{\partial B_\xi} {\rm Tr}'\left[(\boldsymbol{R}^{\alpha}\star \boldsymbol{P}_{\alpha}-2\boldsymbol{L}\right]_\xi\ ,\eea
which vanishes upon imposing
\bea \boldsymbol{P_\a}\vert_{\partial B} ~=~ 0\ ,\eea
and using gauge transitions between charts, acting as follows:
\bea 
\delta_{ \boldsymbol t} \boldsymbol A &=& \boldsymbol D \boldsymbol t^{\, A}-(\boldsymbol t^{\, B}\partial_{ B})\star {\cal F}\ ,\\[5pt]
\delta_{ \boldsymbol t}  \boldsymbol B&=& \boldsymbol D \boldsymbol t^{\, B}-[ \boldsymbol t^{\, A}, B]_\star \ ,\\[5pt]
 \delta_{\boldsymbol t} \boldsymbol U&=&-[ \boldsymbol t^{\, A}, \boldsymbol U]_\star+( \boldsymbol t^{\, B}\partial_{ B})\star( \boldsymbol V\partial_{ B})\star{\cal F}\ ,\\[5pt]
\delta_{ \boldsymbol t} \boldsymbol  V&=& -[ \boldsymbol t^{\, A}, \boldsymbol V]_\star -[ \boldsymbol t^{\, B}, \boldsymbol U ]_\star \ ,
\\[5pt]
 \delta_{\boldsymbol t} \boldsymbol S&=&-[ \boldsymbol t^{\, A}, \boldsymbol S]_\star+( \boldsymbol t^{\, B}\partial_{ B})\star( \boldsymbol T\partial_{ B})\star{\cal F}\ ,\\[5pt]
\delta_{ \boldsymbol t} \boldsymbol  T&=& -[ \boldsymbol t^{\, A}, \boldsymbol T]_\star + \{ \boldsymbol t^{\, B}, \boldsymbol S \}_\star \ ,
\eea
with parameters $(\boldsymbol t^A,\boldsymbol t^B)$ obeying the super-field extension of (\ref{compcondVas}).

\paragraph{Some boundary deformations.}

An example of a set of boundary deformations of minimal bosonic models 
\cite{Sezgin:2011hq} (for the corresponding projection of the off-shell 
system, see \cite{Boulanger:2011dd}) is given by topological vertex operators 
of the form  \cite{Sezgin:2011hq}\footnote{For manifestly Lorentz-covariant 
vertex operators, see \cite{Sezgin:2011hq}.}
\bea 
{\cal V}^{\overrightarrow m}_{[2(m+n)]}&=& {\rm Tr}'\left[
d^4 Z\kappa\star\left( \prod_{i=1}^n \left(R\star E^{2m_i}\right)-
\tfrac{(-1)^n m}{(m+n)}\,E^{\star 2(m+n)}\right)\right]\ ,\label{Vm}\eea
where $\overrightarrow m=(m_1,\dots,m_n)\equiv (m_2,\dots,m_n,m_1)$ with $m_i\geqslant 0$ and $\sum_{i=1}^n m_i=m$ denotes a cyclic order, and 
\bea 
E~:=~\tfrac{1}{2}\,(1-\pi)A_{[1]}\vert_{\mathfrak M}\ ,\qquad 
R~:=~d\Gamma+\Gamma\star \Gamma\ ,\qquad \Gamma~=~\tfrac{1}{2}\,
(1+\pi)A_{[1]}\vert_{\mathfrak M}\ ,
\eea
obeying
\bea 
\nabla E~\approx~0\ ,\qquad R+E\star E~\approx 0\ ,
\eea
with $\nabla=({\rm d}+{\rm ad}_{\Gamma})\vert_{\mathfrak M}$\,. 
It follows that ${\cal V}^{\overrightarrow m}_{[2(m+n)]}$ obeys (\ref{TVO}) (with 
total derivatives on $\mathfrak M$) and that
\bea 
{\cal V}^{\overrightarrow m}_{[2(m+n)]}~\approx~{\cal J}^{\overrightarrow m}_{[2(m+n)]}~:=~\tfrac{(-1)^n n}{(m+n)} \,{\rm Tr}'\left[d^4 Z\kappa\star E^{\star 2(m+n)}\right]\ ,\eea
which is indeed a non-trivial element of the on-shell de Rham cohomology on $
\mathfrak M$ (and hence $\partial \mathfrak M$) assuming a globally-defined 
formulation of fiber-bundle type with structure group containing $\pi$-even but not $\pi$-odd gauge parameters in form degree zero. 
In other words, the insertion of 
${\cal V}^{\overrightarrow m}_{[2(m+n)]}$ at $\partial\mathfrak M$ deforms the 
unbroken phase into a broken phase with smaller structure group and hence additional 
observables; the broken gauge symmetries instead resurface on shell with $\pi$-odd parameters $\xi:=\frac12(1-\pi)\e^A_{[0]}$ forming a section together with the soldering one-form $E$ on a fiber bundle with $\pi$-even structure group in degree zero. 
Indeed, under the gauge transformations $\delta_{\xi}$\,, the variation $\delta_{\xi}{\cal J}^{\overrightarrow m}_{[2(m+n)]}$ consists of total 
derivatives that cancel across chart boundaries provided $(\xi,E)$ forms a section. Clearly, 
the on-shell values of ${\cal V}^{\overrightarrow m}_{[2(m+n)]}$ are non-trivial only on 
submanifolds of $\partial \mathfrak M$ where $E$ is non-degenerate, which is also where 
the parameter $\xi$ can be converted into diffeomorphisms. In other words, perturbing 
the action by $\int_{C}{\cal V}^{\overrightarrow m}_{[2(m+n)]}$ on $2(m+2)$-cycles $C
\subseteq \partial \mathfrak M$, and imposing non-trivial on-shell values for $\int_{C}
{\cal J}^{\overrightarrow m}_{[2(m+n)]}$ leads to a metric phase on $C$\,.

Turning to the total AKSZ master action, it is straightforward to check using the BRST transformations 
\bea s\boldsymbol E~=~\boldsymbol D \boldsymbol E\ ,\qquad 
{\rm s}\boldsymbol \Gamma~=~\boldsymbol R+\boldsymbol E\star \boldsymbol  E\eea
that the BRST transformations of each one of the two terms making up $\boldsymbol V^{\overrightarrow m}_{[2(m+n)]}:={\cal V}^{\overrightarrow m}_{[2(m+n)]}(\boldsymbol Z,{\rm d}\boldsymbol Z)$ transforms into a total derivative such that 
\bea s \boldsymbol V^{\overrightarrow m}_{[2(m+n)]}~=~{\rm d} \boldsymbol W^{\overrightarrow m}_{[2(m+n)]}\ ,\eea
independently of the relative coefficient in ${\cal V}^{\overrightarrow m}_{[2(m+n)]}$, which is instead fixed by demanding the super-field analog of (\ref{TVO}).

\section{Conclusions}\label{sec:conclusions}

In this paper we have taken the first steps of the BV-AKSZ quantization of four-dimensional higher-spin gravity based on the classical action proposed in \cite{Boulanger:2011dd} by constructing the corresponding minimal AKSZ master action obeying the classical BV master equation. We have also given the details of the global formulation within the framework of fiber bundles, which was described only briefly given in \cite{Boulanger:2011dd}.  

Besides the gauge-fixing procedure, which may require non-minimal sectors containing ghost-momenta and Nakanishi-Laudrup auxiliary fields, there are several lines of developments that present themselves at the present stage of which some are: 
\begin{itemize}
\item[(i)] the classification of the bulk Hamiltonians consistent with Vasiliev's theory on the boundary and corresponding globally-defined formulations, that may extend beyond the realm of fiber bundles;
\item[(ii)] the classification of possible deformations of the bulk action, which in general depend on the choice of global formulation in (i);
\item[(iii)] to forgo the associativity of the star-product on the correspondence by considering more general homotopy-associative differential algebras.
\end{itemize}

Finally, it remains an open problem whether contact can be made with the perturbative Fronsdal program. The natural procedure is to add a deformation four-form within a suitable global formulation to be identified as the generating function of holographic correlation functions, possibly in accordance with the various observations and conjectures made in \cite{Sezgin:2002rt,Klebanov:2002ja,Giombi:2010vg,Aharony:2011jz}. 
In the latter respect, the four-form proposed by \cite{Sezgin:2011hq}, that is, the quantity ${\cal V}_{[4]}^{ (2)}$ given in Eq. (\ref{Vm} (for $m=n=1$), which depends only on zero-forms and one-forms on $\partial\mathfrak M$, is an interesting candidate: Assuming that $\hat p=8$ so that ${\rm dim}(\mathfrak M)=5$, and that $\partial \mathfrak M$ is non-compact with non-trivial external states on $\partial^2 \mathfrak M$, it follows that ${\cal V}_{[4]}^{ (2)}$ is non-trivial on-shell (constructed from boundary-to-bulk propagators) and hence a candidate for an on-shell action. Its vertices, on the other hand, cannot be used to close any loops as follows from conservation of form degree on $\mathfrak M$ (bulk vertices of the form ${\rm Tr}' \left[J^r\star U^{\star n}\star V \right]_{{\rm deg}_{\mathfrak M}=5}$ cannot yield correlation functions on $\partial \mathfrak M$ between forms $X^\a|_{\partial \mathfrak M}$ if all degrees $p_\a\leqslant 1$)\,. Hence, it appears treating ${\cal V}_{[4]}^{ (2)}$ as a deformation four-form may give rise to non-trivial tree diagrams and trivial loop corrections, in accordance with the general pattern expected from free conformal field theories.

\paragraph{On-shell equivalence to Fronsdal approach:} 
Concerning the correspondence with the free $O(N)$ vector 
model \cite{Sezgin:2002rt} 
and Gross--Neveu model \cite{Sezgin:2003pt}, 
we make the following observations:
\begin{itemize}
\item for any ${\cal H}(U,V;B)$ and applying perturbation theory in 
which $\int_{\mathfrak M} {\rm Tr}'[dX^\a\star P_\a]$ is treated as the 
kinetic term, it follows from the fact that the vertices in 
${\cal H}(U,V;B)$ are built from exterior (star-) products that boundary 
correlation functions that involve only zero-forms and one-forms are given 
by their semi-classical limits (as vacuum bubbles cancel), \emph{viz.}
\begin{eqnarray} 
&&\langle B_{[0]}(p_1) \cdots B_{[0]}(p_n) A_{[1]}(p_{n+1}) \cdots 
A_{[1]}(p_{n+m})\rangle|_{p_i\in\partial{\mathfrak{M} 
}}\nonumber\\[5pt]&&~=~\langle B_{[0]}(p_1)\rangle \cdots \langle B_{[0]}
(p_n) \rangle\langle A_{[1]}(p_{n+1}) \rangle\cdots \langle A_{[1]}
(p_{n+m})\rangle\ ;
\end{eqnarray}
\item assuming the existence of a perturbative completion 
$\int_{\partial\mathfrak M}{\cal V}_{\rm FV}
(B_{[0]},dB_{[0]};A_{[1]},dA_{[1]})$ of the Fradkin--Vasiliev 
action\footnote{Whether the completion is given in the standard Fronsdal 
formulation or in the frame-like formulation is immaterial as in both cases 
the dynamical field content can be obtained by applying projections to the 
Vasiliev master fields.}, it can be added as a topological vertex operator 
and treated as an interaction, including its kinetic terms;
\item it follows that the expectation value of the Fradkin--Vasiliev action 
is tree-level exact, \emph{i.e.}
\bea Z(\mu)~:=~\left\langle \exp ( \frac{i\mu}\hbar 
\int_{\partial\mathfrak M}{\cal V}_{\rm FV}) \right\rangle~=~\left.\exp 
( \frac{i\mu}\hbar \int_{\partial\mathfrak M}{\cal V}_{\rm FV})
\right|_{B_{[0]}=\langle B_{[0]}\rangle; A_{[1]}=\langle A_{[1]}\rangle}\ ,
\eea
with expectation values $\langle B_{[0]}\rangle$ and 
$\langle A_{[1]}\rangle$ obeying the Vasiliev equations of motion subject 
to boundary conditions at the three-dimensional conformal boundary 
$\bar{\partial}\partial \mathfrak M$ of 
$\partial \mathfrak M$;
\item thus, assuming a suitable topology for $\partial \mathfrak M$ and that 
$\langle B_{[0]}\rangle$ and $\langle A_{[1]}\rangle$ are asymptotic to 
$AdS_4$, hence built from the boundary data using boundary-to-bulk 
propagators, we expect that $Z(\mu)$ with $\mu N=\hbar$ is equal to the 
generating functional of the free $O(N)$ model in the case of the Type A 
model with scalar field obeying $\Delta=1$ boundary conditions, and to the 
generating functional of the free Gross--Neveu model (with $N$ free 
fermions) in the case of the Type B model with scalar field obeying 
$\Delta=2$ boundary conditions.
\end{itemize}
We wish to stress the fact that both of the latter higher-spin gravity 
models are manifestly tree-level unitary: by the very nature of the 
perturbative treatment of the Poisson sigma models 
(with kinetic $P dX$-terms on $\mathfrak M$), the partition function 
$Z(\mu)$ is completely free from loop corrections in the Fradkin--Vasiliev 
sector, in perfect agreement with free three-dimensional CFTs. In other 
words, $Z(\mu)$ is given by the sum of tree Witten-diagrams in $AdS_4$ with 
external boundary-to-bulk and internal bulk-to-bulk Green's functions 
arising as the result of solving classical equations of motion subject to 
boundary sources (and not of performing any Gaussian integrals starting from 
the Fronsdal kinetic terms in the Fradkin--Vasiliev action).

In the case of the strongly-coupled fixed points of the $O(N)$ vector model 
\cite{Klebanov:2002ja} and the Gross--Neveu model \cite{Sezgin:2003pt}, 
reached by suitable double-trace deformations, the Fradkin--Vasiliev action 
needs to be modified with a Gibbons--Hawking term
\bea \int_{\bar{\partial}\partial \mathfrak M}
{\cal V}_{\rm GH}~=~\int_{\bar{\partial}\partial \mathfrak M} \phi \,
\partial_n \phi  +\cdots\ ,\eea
where the $\cdots$ contain a non-linear completion achieving higher-spin 
gauge invariance.

In the standard perturbative approach, in which the kinetic terms are taken 
from $\int_{\partial\mathfrak M}{\cal V}_{\rm FV}\,$, this modification 
induces a shift in the scalar two-point function $G_{\Delta=1}$ as follows (for a recent treatment, see \cite{Giombi:2011ya}):
\bea G_{\Delta=1}(p;r,r')+|p| K_{\Delta=1}(p;r)
K_{\Delta=1}(p;r')~\equiv~G_{\Delta=2}(p;r,r')\ .\eea

In the Poisson sigma model, on the other hand, the Gibbons--Hawking 
modification is instead treated as an additional vertex. As a result, pairs 
of external scalar legs of the tree diagrams are sewn together leading to 
additional scalar loops that are restricted in the configuration space as to 
touch the boundary. Likewise, the non-linear completion of 
$\int_{\bar{\partial}\partial \mathfrak M}{\cal V}_{\rm GH}$ may induce loop-corrections 
involving higher-spin fields running in similar boundary loops.

\paragraph*{Acknowledgements}

It is a pleasure to thank C. Iazeolla and A. Sagnotti for useful 
interactions and early collaboration.
N.B. and P.S. thank the Scuola Normale Superiore, Pisa, 
where this work was initiated.
We thank G. Barnich, M. Grigoriev, S. Lyakhovich, E. Sezgin, 
E.D. Skvortsov, T. Strobl, 
M. Valenzuela, M.~Vasiliev and R. Zucchini for 
discussions. N.B. and P.S. thank the Erwin Schr\"odinger Institute 
for kind hospitality during the Higher Spin Gravity Workshop organized there.
The work of N.B. was supported in parts by an ARC contract No. 
AUWB-2010-10/15-UMONS-1.

\section*{Appendix}

\paragraph{Star-vector fields} 

A graded-associative quasi-free differential algebra 
on a non-commutative base manifold $\mathfrak B\,$ consists of local 
representatives ${\mathfrak{R}}_{\xi}$ 
($\xi$ labels charts $B_\xi \subset  B$)
generated by sets $\{Z^i_\xi\}_{i\in{\cal S}}$ 
of locally-defined differential forms subject to generalized curvature constraints
\begin{eqnarray}
 {\cal R}^i_\xi~:=~{\rm d} Z^i_\xi+{\cal Q}^i(Z_\xi , J)~\approx~0\ ,
\end{eqnarray}
where $\overrightarrow{\cal Q}:={\cal Q}^i\,\partial_i\,$
(with $\partial_i\equiv \overrightarrow\partial_i$) is a composite 
$\star$-vector field
of total degree one subject to the Cartan integrability condition
\begin{eqnarray}
\overrightarrow{\cal Q}\star {\cal Q}^i~\equiv~0\ .\label{CIcond}
\end{eqnarray}

A composite $\star$-vector field $\overrightarrow{\cal X}$ 
(see Appendix B of 
\cite{Boulanger:2011dd} for more details) is a graded inner derivation
 of the graded associative $\star$-product algebra
 ${\mathfrak{R}}:={\rm Env}[Z^i]\otimes \mathfrak J$
 where $\mathfrak J$ is a space of central and $\rm d$-closed elements (including the
 identity), \emph{i.e.} if ${\cal F},{\cal F}'\in {\mathfrak{R}}$ then
 \begin{eqnarray}
 \overrightarrow{\cal X}\star ({\cal F}\star {\cal F}')~=~(\overrightarrow{\cal X}\star {\cal F})\star
 {\cal F}'+(-1)^{{\rm deg}(\overrightarrow{\cal X}){\rm deg}({\cal F})}{\cal F}\star
 (\overrightarrow{\cal X}\star {\cal F}')\ , \label{Leibnitz}
 \end{eqnarray}
 provided that $\overrightarrow{\cal X}$ and ${\cal F}$ have fixed degrees.
 In components, one writes $\overrightarrow{\cal X}:={\cal X}^{ i}( Z^{ j}) \partial_{ i}$
 where
 ${\cal X}^{ i}:=\overrightarrow{\cal X}\star Z^{ i}\,$.
 The graded bracket between two composite $\star$-vector fields is defined by
 \begin{eqnarray}
 [\overrightarrow{\cal X},\overrightarrow{\cal X}']_{\star}\star {\cal F}
~:=~\overrightarrow{\cal X}\star(\overrightarrow{\cal X}'\star
 {\cal F})-(-1)^{{\rm deg}(\overrightarrow{\cal X}){\rm deg}
(\overrightarrow{\cal X}')}\overrightarrow{\cal X}'\star
 (\overrightarrow{\cal X}\star{\cal F})\ ,
 \end{eqnarray}
 is a degree-preserving graded Lie bracket, \emph{i.e.}
 $[\overrightarrow{\cal X},\overrightarrow{\cal X}']_{\star}$ is a graded inner derivation  
obeying the graded
 Jacobi identity
 $[[\overrightarrow{\cal X},\overrightarrow{\cal X}']_{\star},\overrightarrow{\cal X}'']_\star
+\mbox{graded cyclic}\equiv 0\,$.
In  components, one has
 \begin{eqnarray}
 [\overrightarrow{\cal X},\overrightarrow{\cal X}']_{\star} ~=~ 
\left(\overrightarrow{\cal X}\star {\cal X}^{\prime i}
 - (-1)^{\overrightarrow{\rm deg}({\cal X}){\rm deg}(\overrightarrow{\cal X}')}\overrightarrow{\cal X}'
\star {\cal X}^i\right)\partial_i\ .
 \end{eqnarray}
The Cartan integrability condition (\ref{CIcond}), that can be rewritten
$[\overrightarrow{\cal Q},\overrightarrow{\cal Q}]_\star\equiv 0\,$, amounts to that
$\overrightarrow{\cal Q}$ is a nilpotent composite $\star$-vector field of degree one.
This condition ensures that the generalized curvature constraints
${\cal R}^i_\xi\approx 0$ are compatible with ${\rm d}^2\equiv 0$ without further algebraic
constraints on the generating elements $Z^i_\xi\,$.
One can also show \cite{Boulanger:2011dd} 
that the nilpotency of $\overrightarrow{\cal Q}$ is separately equivalent to that
the generalized curvatures ${\cal R}^i$ obey the generalized Bianchi identities
\begin{eqnarray}
{\rm d}{\cal R}^i-\overrightarrow{\cal R}\star {\cal Q}^i~\equiv~0\ ,
\qquad{\rm{where}} \qquad \overrightarrow{\cal R}:={\cal R}^i \,\partial_i \ ,
\end{eqnarray}
and transform into each other under the following Cartan gauge transformations
\begin{eqnarray}
 \delta_{\varepsilon}Z^i
~\equiv ~{\cal T}_{\varepsilon}^i
 ~:=~  d\varepsilon^i - \overrightarrow\varepsilon \star {\cal Q}^i \ ,
\qquad{\rm{where}} \qquad \overrightarrow\varepsilon:=\varepsilon^i \,\partial_i  \
\end{eqnarray}
and where $\varepsilon^i$ is considered infinitesimal
and independent of $Z^i\,$, \emph{viz.}
\begin{eqnarray}
 \delta_{\varepsilon}{\cal R}^i~=~ - \overrightarrow{\cal R}\star
 \left( (\overrightarrow\varepsilon\star {\cal Q}^i)\right).
\end{eqnarray}

\paragraph{Functional derivative on commutative manifold:}
\noindent  We define the variational functional left derivative
$\d_{f(p)}F[f]\equiv\frac{\delta^L}{\delta f(p)} F[f]$ at $p\in B$ 
of a functional $F[f]$ with respect to a differential form $f$ via the relation
\begin{equation}
 \int_{p\in B} \delta f(p) \;\d_{f(p)} F[f]~=~F[f+\d f]-F[f]+O((\d f)^2) \ .
\label{funcderivcomm}
\end{equation}
We assign a total degree and a Grassmann parity, respectively, to variables, operations and maps as follows:
\bea |\cdot|&:=& {\rm deg}(\cdot) + {\rm gh} (\cdot)\ ,\qquad 
{\rm Gr}(\cdot)~=~|\cdot|\quad{\rm mod}~2\ ,
\eea
which implies that the total exterior derivative ${\rm d}$ anti-commutes with the BRST operator. 
We refer to a functional $F[f]$ as being \emph{ultra-local} if $F[f]=L(f,{\rm d}f)$ 
where $L$ is an algebraic function of $f$ and ${\rm d}f\,$, 
and as being \emph{local} if $F[f]=\int_{B} {\cal L}(f,{\rm d}f)$ where ${\cal L}$ is ultra-local. 
We refer to a functional as being \emph{intrinsically defined} on $B$ 
if it does not refer to any auxiliary frame on $B\,$. 
The functional derivatives of local functionals are intrinsically defined and ultra-local, 
\emph{viz.}
\bea 
\d_{f(p)} \int_{B} {\cal L}(f,{\rm d}f)&=& \left(\partial_f {\cal L}-(-1)^{|f|} {\rm d}
(\partial_{{\rm d}f} {\cal L})\right)(p) ~ \stackrel{def.}{=} ~ \frac{\delta {\cal L}(f,{\rm d}f)}{\delta f}(p)\; \ ,
\label{functionderiv}
\eea 
where throughout the paper all the derivatives are left-derivatives, so that
$\partial_f {\cal L} =\frac{\partial^{L}}{\partial f}{\cal L} \,$ and 
$\partial_{{\rm d}f} {\cal L} = \frac{\partial^{L}}{\partial {\rm d}f}{\cal L} \,$.
The functional derivatives of ultra-local functionals are given by
\bea 
\d_{f(p)} \left(L(f,{\rm d}f)(p')\right) &=& [\d_{f(p)} f(p')] \left(\partial_f L\right)(p') 
+ (-1)^{\hat p+1+|f|}\left({\rm d}_{p'}[\d_{f(p)} f(p')]\right)(\partial_{{\rm d}f}L)(p')\ ,\qquad
\eea
and refers to an auxiliary frame $h^A$ via the distribution (taking $f$ to be a 
$q$-form):
\bea 
\frac{\delta f(p)}{\delta f(p')}~\equiv~ \delta_{f(p)} f(p') &=& (-1)^{\hat p |f|+{\rm gh}(f)} ~h^{A[\hat p +1 - q]}(p)~h^{B[q]}(p')~
\e_{A[\hat p+1-q]B[q]}~\d(p,p')  \ ,
\eea
where the Dirac function is the zero-form defined by
\bea 
\int_{p\in B} h(p) \varphi(p)\delta(p,p')& = &\varphi(p')\ ,\qquad \varphi\ \in\ \O^{[0]}(B)\ ,
\eea
where we use the definitions and conventions
\be 
h^{A[n]}~=~\frac1{n!} h^{A_1}\cdots h^{A_n}\ ,\qquad h~=~h^{A[\hat p+1]}\e_{A[\hat p+1]}\ ,
\ee
\be
\e^{A[n]C[\hat p+1-n]}\e_{B[n]C[\hat p+1-n]}\ =\ (-1)^{\eta_{AB}} n!(\hat p+1-n)!\d^{A[n]}_{B[n]}\ .
\ee
Then, the functional derivative of an ultra-local functional $F(f,{\rm d}f)$ is such that one has
\begin{eqnarray}
\int_{p\in B} \delta f(p)\;\left[\d_{f(p)} \left(F(f,{\rm d}f)(p')\right)\right] 
&=&  \delta f(p') \;\frac{\delta F}{\delta f}(p')
+\;{\rm d}_{p'}\left[ \delta f(p')\; \left(\partial_{{\rm d}f} { F}\right)(p') \right]
\end{eqnarray}
using the notation and definition of (\ref{functionderiv}). 
Therefore, expanding the total derivative on the right-hand side of the above equation, one has 
\begin{equation}
 F(f+\delta f,{\rm d}(f+\delta f))(p')-F(f,{\rm d}f)(p')=
({\rm d}_{p'} \delta f(p'))\, \partial_{{\rm d}f} F (p') + 
 \delta f(p') \,\partial_{f}F(p')\;.
\end{equation}

\paragraph{Functional variations in the non-commutative case}

In the case of a non-commutative graded manifold one defines the functional variation 
$\frac{\delta F}{\delta Z^i}$ of a functional $F[Z]$ by 
\begin{equation}
 F[Z +\delta F]-F[Z] ~=~ \delta F ~=~ \int_{p\in B}\left(\delta Z^i(p) \star \frac{\delta F[Z]}{\delta Z^i(p)}\right) + 
{\cal O}((\delta Z)^2)\quad .
\end{equation}

Starting from the functional 
$F[Z]=\int_B L_\star(Z,{\rm d}Z)\,$ where 
$L_\star(Z,{\rm d}Z)$ is a star-function of $(Z,{\rm d}Z)\,$, 
one has  
\begin{equation}
\frac{\delta F[Z]}{\delta Z^i(p)}= {\partial_i}^{cycl} L_\star(p) 
 - (-1)^i\,{\rm d} (\partial^{cycl}_{{\rm d}Z^i}L_\star)(p)=:
 \frac{\delta L_\star(Z,{\rm d}Z)}{\delta Z^i}(p)
\end{equation}
where, for  
$P_\star(Z)=f_{i_1,\ldots ,i_n}\;Z^{i_1}\star \ldots\star 
 Z^{i_n}\equiv(-1)^{i_1(i_2+\ldots+i_n)}
 f_{i_2,\ldots ,i_n, i_1}\;Z^{i_1}\star \ldots\star Z^{i_1}\,$, 
the \emph{cyclic derivative} 
\begin{eqnarray}
&\partial^{cycl}_i P_\star(Z) = n\;
f_{i, i_2,\ldots ,i_n}\;Z^{i_2}\star \ldots\star Z^{i_n}\quad .&
\end{eqnarray}
One then defines
\begin{eqnarray}
\frac{\delta }{\delta Z^i(p)}\left[L_{\star}(Z,{\rm d}Z)(p')\right] =  
\frac{\delta Z^j(p')}{\delta Z^i(p)}\;\star 
\frac{\partial^{cycl} L_{\star}}{\partial Z^j}\;(p') + 
(-1)^{\hat{p}+i+1}
 \left( {\rm{d}}_{p'}\;\frac{\delta Z^j(p')}{\delta Z^i(p)}\;\right)
\star \frac{\partial^{cycl} 
L_{\star}}{\partial {\rm{d}} Z^j}\;(p')
\end{eqnarray}
where $\frac{\delta Z^j(p')}{\delta Z^i(p)}$ has total degree 
$j-i-\hat{p}-1$ and is such that 
\begin{eqnarray}
\int_{p\in\mathfrak{M}}{{\rm Tr}}\left[ \delta Z^i(p)\star
 \frac{\delta Z^j(p')}{\delta Z^i(p)} \star 
 \frac{\partial^{cycl} L_\star}{\partial Z^j}(p')\right]=\delta Z^i(p')\star
 \frac{\partial^{cycl} L_\star}{\partial Z^i}(p')\quad. 
\end{eqnarray}
As a result, the action of 
$\delta = \int_{p\in B} \delta Z^ i(p)\star\frac{\delta}{\delta Z^i(p)}$
on the ultra-local functional $L_\star(Z,{\rm d}Z)(p')$ yields
\begin{eqnarray}
\delta L_\star(Z,{\rm d}Z)(p') &=& \delta Z^i(p')\star 
\frac{\delta L_\star}{\delta Z^i}(p')+{\rm d}_{p'}\left[
\delta Z^i(p')\star 
\frac{\partial^{\rm cycl} L_\star}{\partial {\rm d}Z^i}(p')
\right]\;
\nonumber \\
&=& \delta Z^i(p')\star\frac{\partial^{\rm cycl} L_\star}{\partial Z^i}(p')
 + \delta ({\rm d}Z^i)(p')\star\frac{\partial^{\rm cycl} L_\star}{\partial {\rm d}Z^i}(p')\quad,
\end{eqnarray}
as it should. 


\providecommand{\href}[2]{#2}\begingroup\raggedright\endgroup


\begin{thebibliography}{10}

\bibitem{Boulanger:2011dd}
N.~Boulanger and P.~Sundell, ``{An action principle for Vasiliev's
  four-dimensional higher-spin gravity},''
  \href{http://dx.doi.org/10.1088/1751-8113/44/49/495402}{{\em J.Phys.A} {\bf
  A44} (2011)  495402}, \href{http://arxiv.org/abs/1102.2219}{{\tt
  arXiv:1102.2219 [hep-th]}}.

\bibitem{Vasiliev:1990en}
M.~A. Vasiliev, ``{Consistent equation for interacting gauge fields of all
  spins in (3+1)-dimensions},''
  \href{http://dx.doi.org/10.1016/0370-2693(90)91400-6}{{\em Phys. Lett.} {\bf
  B243} (1990)  378--382}.
\url{http://dx.doi.org/10.1016/0370-2693(90)91400-6}.

\bibitem{Vasiliev:1990vu}
M.~A. Vasiliev, ``{Properties of equations of motion of interacting gauge
  fields of all spins in (3+1)-dimensions},''
\href{http://dx.doi.org/10.1088/0264-9381/8/7/014}{{\em Class. Quant. Grav.}
  {\bf 8} (1991)  1387--1417}.

\bibitem{Vasiliev:1992av}
M.~A. Vasiliev, ``{More on equations of motion for interacting massless fields
  of all spins in (3+1)-dimensions},''
\href{http://dx.doi.org/10.1016/0370-2693(92)91457-K}{{\em Phys. Lett.} {\bf
  B285} (1992)  225--234}.

\bibitem{Sezgin:2002ru}
E.~Sezgin and P.~Sundell, ``{Analysis of higher spin field equations in four
  dimensions},'' {\em JHEP} {\bf 07} (2002)  055,
\href{http://arxiv.org/abs/hep-th/0205132}{{\tt arXiv:hep-th/0205132}}.

\bibitem{Vasiliev:1988xc}
M.~A. Vasiliev, ``{Equations of motion of interacting massless fields of all
  spins as a free differential algebra},''
\href{http://dx.doi.org/10.1016/0370-2693(88)91179-3}{{\em Phys. Lett.} {\bf
  B209} (1988)  491--497}.

\bibitem{Vasiliev:1988sa}
M.~A. Vasiliev, ``{Consistent equations for interacting massless fields of all
  spins in the first order in curvatures},''
\href{http://dx.doi.org/10.1016/0003-4916(89)90261-3}{{\em Annals Phys.} {\bf
  190} (1989)  59--106}.

\bibitem{Vasiliev:1992gr}
M.~A. Vasiliev, ``{Unfolded representation for relativistic equations in (2+1)
  anti-De Sitter space},''
{\em Class. Quant. Grav.} {\bf 11} (1994)  649--664.

\bibitem{Vasiliev:2005zu}
M.~A. Vasiliev, ``{Actions, charges and off-shell fields in the unfolded
  dynamics approach},'' {\em Int. J. Geom. Meth. Mod. Phys.} {\bf 3} (2006)
  37--80,
\href{http://arxiv.org/abs/hep-th/0504090}{{\tt arXiv:hep-th/0504090}}.

\bibitem{Ikeda:1993fh}
N.~Ikeda, ``{Two-dimensional gravity and nonlinear gauge theory},''
  \href{http://dx.doi.org/10.1006/aphy.1994.1104}{{\em Ann. Phys.} {\bf 235}
  (1994)  435--464},
\href{http://arxiv.org/abs/hep-th/9312059}{{\tt arXiv:hep-th/9312059}}.

\bibitem{Schaller:1994es}
P.~Schaller and T.~Strobl, ``{Poisson structure induced (topological) field
  theories},'' \href{http://dx.doi.org/10.1142/S0217732394002951}{{\em
  Mod.Phys.Lett.} {\bf A9} (1994)  3129--3136},
\href{http://arxiv.org/abs/hep-th/9405110}{{\tt arXiv:hep-th/9405110
  [hep-th]}}.

\bibitem{Cattaneo:2001ys}
A.~S. Cattaneo and G.~Felder, ``{On the AKSZ formulation of the Poisson sigma
  model},'' \href{http://dx.doi.org/10.1023/A:1010963926853}{{\em Lett. Math.
  Phys.} {\bf 56} (2001)  163--179},
  \href{http://arxiv.org/abs/math/0102108}{{\tt arXiv:math/0102108}}.
\url{http://dx.doi.org/10.1023/A:1010963926853}.

\bibitem{Batalin:1981jr}
I.~Batalin and G.~Vilkovisky, ``{Gauge Algebra and Quantization},''
  \href{http://dx.doi.org/10.1016/0370-2693(81)90205-7}{{\em Phys.Lett.} {\bf
  B102} (1981)  27--31}.

\bibitem{Batalin:1984jr}
I.~Batalin and G.~Vilkovisky, ``{Quantization of Gauge Theories with Linearly
  Dependent Generators},'' \href{http://dx.doi.org/10.1103/PhysRevD.28.2567,
  10.1103/PhysRevD.30.508}{{\em Phys.Rev.} {\bf D28} (1983)  2567--2582}.

\bibitem{Alexandrov:1995kv}
M.~Alexandrov, M.~Kontsevich, A.~Schwartz, and O.~Zaboronsky, ``{The Geometry
  of the master equation and topological quantum field theory},''
  \href{http://dx.doi.org/10.1142/S0217751X97001031}{{\em Int. J. Mod. Phys.}
  {\bf A12} (1997)  1405--1430},
  \href{http://arxiv.org/abs/hep-th/9502010}{{\tt arXiv:hep-th/9502010}}.
\url{http://dx.doi.org/10.1142/S0217751X97001031}.

\bibitem{Cattaneo:1999fm}
A.~S. Cattaneo and G.~Felder, ``{A path integral approach to the Kontsevich
  quantization formula},'' \href{http://dx.doi.org/10.1007/s002200000229}{{\em
  Commun. Math. Phys.} {\bf 212} (2000)  591--611},
  \href{http://arxiv.org/abs/math/9902090}{{\tt arXiv:math/9902090}}.
\url{http://dx.doi.org/10.1007/s002200000229}.

\bibitem{Cattaneo:2001bp}
A.~S. Cattaneo and G.~Felder, ``{Poisson sigma models and deformation
  quantization},'' \href{http://dx.doi.org/10.1142/S0217732301003255}{{\em Mod.
  Phys. Lett.} {\bf A16} (2001)  179--190},
  \href{http://arxiv.org/abs/hep-th/0102208}{{\tt arXiv:hep-th/0102208}}.
\url{http://dx.doi.org/10.1142/S0217732301003255}.

\bibitem{Kontsevich:1997vb}
M.~Kontsevich, ``{Deformation quantization of Poisson manifolds. 1.},''
  \href{http://dx.doi.org/10.1023/B:MATH.0000027508.00421.bf}{{\em
  Lett.Math.Phys.} {\bf 66} (2003)  157--216},
  \href{http://arxiv.org/abs/q-alg/9709040}{{\tt arXiv:q-alg/9709040 [q-alg]}}.

\bibitem{Birmingham:1991ty}
D.~Birmingham, M.~Blau, M.~Rakowski, and G.~Thompson, ``{Topological field
  theory},''
\href{http://dx.doi.org/10.1016/0370-1573(91)90117-5}{{\em Phys.Rept.} {\bf
  209} (1991)  129--340}.

\bibitem{Blasi:2005vf}
A.~Blasi, N.~Maggiore, and M.~Montobbio, ``{Noncommutative two dimensional BF
  model},'' \href{http://dx.doi.org/10.1016/j.nuclphysb.2006.01.028}{{\em
  Nucl.Phys.} {\bf B740} (2006)  281--296},
\href{http://arxiv.org/abs/hep-th/0512006}{{\tt arXiv:hep-th/0512006
  [hep-th]}}.

\bibitem{Vilar:2007iu}
L.~Vilar, O.~Ventura, R.~Amaral, V.~Lemes, and L.~Buffon, ``{Seiberg-Witten map
  for the 4D noncommutative BF theory},''
  \href{http://dx.doi.org/10.1088/1751-8113/41/42/425203}{{\em J.Phys.A} {\bf
  A41} (2008)  425203},
\href{http://arxiv.org/abs/0710.3954}{{\tt arXiv:0710.3954 [hep-th]}}.

\bibitem{Grigoriev:1999qz}
M.~Grigoriev and P.~Damgaard, ``{Superfield BRST charge and the master
  action},'' \href{http://dx.doi.org/10.1016/S0370-2693(00)00050-2}{{\em
  Phys.Lett.} {\bf B474} (2000)  323--330},
  \href{http://arxiv.org/abs/hep-th/9911092}{{\tt arXiv:hep-th/9911092
  [hep-th]}}.

\bibitem{Park:2000au}
J.-S. Park, ``{Topological open p-branes},''
\href{http://arxiv.org/abs/hep-th/0012141}{{\tt arXiv:hep-th/0012141}}.

\bibitem{Hofman:2002rv}
C.~Hofman and J.-S. Park, ``{Topological open membranes},''
\href{http://arxiv.org/abs/hep-th/0209148}{{\tt arXiv:hep-th/0209148}}.

\bibitem{Ikeda:2000yq}
N.~Ikeda, ``{A deformation of three dimensional BF theory},'' {\em JHEP} {\bf
  11} (2000)  009,
\href{http://arxiv.org/abs/hep-th/0010096}{{\tt arXiv:hep-th/0010096}}.

\bibitem{Ikeda:2001fq}
N.~Ikeda, ``{Deformation of BF theories, topological open membrane and a
  generalization of the star deformation},'' {\em JHEP} {\bf 07} (2001)  037,
\href{http://arxiv.org/abs/hep-th/0105286}{{\tt arXiv:hep-th/0105286}}.

\bibitem{Roytenberg:2002nu}
D.~Roytenberg, ``{On the structure of graded symplectic supermanifolds and
  Courant algebroids},'' in {\em Quantization, {P}oisson brackets and beyond
  ({M}anchester, 2001)}, vol.~315 of {\em Contemp. Math.}, pp.~169--185.
\newblock Amer. Math. Soc., Providence, RI, 2002.
\newblock
\href{http://arxiv.org/abs/math/0203110}{{\tt arXiv:math/0203110}}.
\newblock

\bibitem{Hofman:2002jz}
C.~Hofman and J.-S. Park, ``{BV quantization of topological open membranes},''
  \href{http://dx.doi.org/10.1007/s00220-004-1106-7}{{\em Commun. Math. Phys.}
  {\bf 249} (2004)  249--271},
\href{http://arxiv.org/abs/hep-th/0209214}{{\tt arXiv:hep-th/0209214}}.

\bibitem{Ikeda:2002wh}
N.~Ikeda, ``{Chern-Simons gauge theory coupled with BF theory},''
  \href{http://dx.doi.org/10.1142/S0217751X03015155}{{\em Int. J. Mod. Phys.}
  {\bf A18} (2003)  2689--2702},
\href{http://arxiv.org/abs/hep-th/0203043}{{\tt arXiv:hep-th/0203043}}.

\bibitem{Ikeda:2006wd}
N.~Ikeda, ``{Deformation of Batalin-Vilkovisky Structures},''
\href{http://arxiv.org/abs/math/0604157}{{\tt arXiv:math/0604157}}.

\bibitem{Roytenberg:2006qz}
D.~Roytenberg, ``{AKSZ-BV formalism and Courant algebroid-induced topological
  field theories},'' \href{http://dx.doi.org/10.1007/s11005-006-0134-y}{{\em
  Lett. Math. Phys.} {\bf 79} (2007)  143--159},
\href{http://arxiv.org/abs/hep-th/0608150}{{\tt arXiv:hep-th/0608150}}.

\bibitem{Barnich:2009jy}
G.~Barnich and M.~Grigoriev, ``{A Poincare lemma for sigma models of AKSZ
  type},'' \href{http://dx.doi.org/10.1016/j.geomphys.2010.11.014}{{\em
  J.Geom.Phys.} {\bf 61} (2011)  663--674},
\href{http://arxiv.org/abs/0905.0547}{{\tt arXiv:0905.0547 [math-ph]}}.

\bibitem{Ikeda:2012pv}
N.~Ikeda, ``{Lectures on AKSZ Topological Field Theories for Physicists},''
\href{http://arxiv.org/abs/1204.3714}{{\tt arXiv:1204.3714 [hep-th]}}.

\bibitem{Barnich:2004cr}
G.~Barnich, M.~Grigoriev, A.~Semikhatov, and I.~Tipunin, ``{Parent field theory
  and unfolding in BRST first-quantized terms},''
  \href{http://dx.doi.org/10.1007/s00220-005-1408-4}{{\em Commun.Math.Phys.}
  {\bf 260} (2005)  147--181}, \href{http://arxiv.org/abs/hep-th/0406192}{{\tt
  arXiv:hep-th/0406192 [hep-th]}}.

\bibitem{Barnich:2005ru}
G.~Barnich and M.~Grigoriev, ``{BRST extension of the non-linear unfolded
  formalism},''
\href{http://arxiv.org/abs/hep-th/0504119}{{\tt arXiv:hep-th/0504119}}.

\bibitem{Grigoriev:2006tt}
M.~Grigoriev, ``{Off-shell gauge fields from BRST quantization},''
\href{http://arxiv.org/abs/hep-th/0605089}{{\tt arXiv:hep-th/0605089}}.

\bibitem{Barnich:2010sw}
G.~Barnich and M.~Grigoriev, ``{First order parent formulation for generic
  gauge field theories},''
  \href{http://dx.doi.org/10.1007/JHEP01(2011)122}{{\em JHEP} {\bf 1101} (2011)
   122},
\href{http://arxiv.org/abs/1009.0190}{{\tt arXiv:1009.0190 [hep-th]}}.

\bibitem{Grigoriev:2010ic}
M.~Grigoriev, ``{Parent formulation at the Lagrangian level},''
  \href{http://dx.doi.org/10.1007/JHEP07(2011)061}{{\em JHEP} {\bf 1107} (2011)
   061},
\href{http://arxiv.org/abs/1012.1903}{{\tt arXiv:1012.1903 [hep-th]}}.

\bibitem{Kaparulin:2011zz}
D.~Kaparulin, S.~Lyakhovich, and A.~Sharapov, ``{On Lagrange structure of
  unfolded field theory},''
  \href{http://dx.doi.org/10.1142/S0217751X11052840}{{\em Int.J.Mod.Phys.} {\bf
  A26} (2011)  1347--1362},
\href{http://arxiv.org/abs/1012.2567}{{\tt arXiv:1012.2567 [hep-th]}}.

\bibitem{Kazinski:2005eb}
P.~O. Kazinski, S.~L. Lyakhovich, and A.~A. Sharapov, ``{Lagrange structure and
  quantization},'' \href{http://dx.doi.org/10.1088/1126-6708/2005/07/076}{{\em
  JHEP} {\bf 07} (2005)  076}, \href{http://arxiv.org/abs/hep-th/0506093}{{\tt
  arXiv:hep-th/0506093}}.
\url{http://dx.doi.org/10.1088/1126-6708/2005/07/076}.

\bibitem{Lyakhovich:2006sc}
S.~L. Lyakhovich and A.~A. Sharapov, ``{Quantizing non-Lagrangian gauge
  theories: An augmentation method},''
  \href{http://dx.doi.org/10.1088/1126-6708/2007/01/047}{{\em JHEP} {\bf 01}
  (2007)  047}, \href{http://arxiv.org/abs/hep-th/0612086}{{\tt
  arXiv:hep-th/0612086}}.
\url{http://dx.doi.org/10.1088/1126-6708/2007/01/047}.

\bibitem{Zucchini:2008hn}
R.~Zucchini, ``{The Lie algebroid Poisson sigma model},''
  \href{http://dx.doi.org/10.1088/1126-6708/2008/12/062}{{\em JHEP} {\bf 0812}
  (2008)  062}, \href{http://arxiv.org/abs/0810.3300}{{\tt arXiv:0810.3300
  [math-ph]}}.

\bibitem{Kotov:2007nr}
A.~Kotov and T.~Strobl, ``{Characteristic classes associated to Q-bundles},''
\href{http://arxiv.org/abs/0711.4106}{{\tt arXiv:0711.4106 [math.DG]}}.

\bibitem{Kotov:2010wr}
A.~Kotov and T.~Strobl, ``{Generalizing Geometry - Algebroids and Sigma
  Models},''
\href{http://arxiv.org/abs/1004.0632}{{\tt arXiv:1004.0632 [hep-th]}}.

\bibitem{Becchi:1974xu}
C.~Becchi, A.~Rouet, and R.~Stora, ``{The Abelian Higgs-Kibble Model. Unitarity
  of the S Operator},''
  \href{http://dx.doi.org/10.1016/0370-2693(74)90058-6}{{\em Phys.Lett.} {\bf
  B52} (1974)  344}.

\bibitem{Becchi:1974md}
C.~Becchi, A.~Rouet, and R.~Stora, ``{Renormalization of the Abelian
  Higgs-Kibble Model},'' \href{http://dx.doi.org/10.1007/BF01614158}{{\em
  Commun.Math.Phys.} {\bf 42} (1975)  127--162}.

\bibitem{Becchi:1975nq}
C.~Becchi, A.~Rouet, and R.~Stora, ``{Renormalization of Gauge Theories},''
  \href{http://dx.doi.org/10.1016/0003-4916(76)90156-1}{{\em Annals Phys.} {\bf
  98} (1976)  287--321}.

\bibitem{Tyutin:1975qk}
I.~Tyutin, ``{Gauge Invariance in Field Theory and Statistical Physics in
  Operator Formalism},'' \href{http://arxiv.org/abs/0812.0580}{{\tt
  arXiv:0812.0580 [hep-th]}}.

\bibitem{Colombo:2010fu}
N.~Colombo and P.~Sundell, ``{Twistor space observables and quasi-amplitudes in
  4D higher spin gravity},''
  \href{http://dx.doi.org/10.1007/JHEP11(2011)042}{{\em JHEP} {\bf 1111} (2011)
   042},
\href{http://arxiv.org/abs/1012.0813}{{\tt arXiv:1012.0813 [hep-th]}}.

\bibitem{Iazeolla:2011cb}
C.~Iazeolla and P.~Sundell, ``{Families of exact solutions to Vasiliev's 4D
  equations with spherical, cylindrical and biaxial symmetry},''
  \href{http://dx.doi.org/10.1007/JHEP12(2011)084}{{\em JHEP} {\bf 1112} (2011)
   084},
\href{http://arxiv.org/abs/1107.1217}{{\tt arXiv:1107.1217 [hep-th]}}.

\bibitem{Sezgin:2011hq}
E.~Sezgin and P.~Sundell, ``{Geometry and Observables in Vasiliev's Higher Spin
  Gravity},'' \href{http://dx.doi.org/10.1007/JHEP07(2012)121}{{\em JHEP} {\bf
  1207} (2012)  121},
\href{http://arxiv.org/abs/1103.2360}{{\tt arXiv:1103.2360 [hep-th]}}.

\bibitem{Perez:2012cf}
A.~Perez, D.~Tempo, and R.~Troncoso, ``{Higher spin gravity in 3D: black holes,
  global charges and thermodynamics},''
\href{http://arxiv.org/abs/1207.2844}{{\tt arXiv:1207.2844 [hep-th]}}.

\bibitem{Barnich:2000zw}
G.~Barnich, F.~Brandt, and M.~Henneaux, ``{Local BRST cohomology in gauge
  theories},'' \href{http://dx.doi.org/10.1016/S0370-1573(00)00049-1}{{\em
  Phys.Rept.} {\bf 338} (2000)  439--569},
\href{http://arxiv.org/abs/hep-th/0002245}{{\tt arXiv:hep-th/0002245
  [hep-th]}}.

\bibitem{Henneaux:1992ig}
M.~Henneaux and C.~Teitelboim, ``{Quantization of gauge systems},''. Princeton,
  USA: Univ. Pr. (1992) 520 p.

\bibitem{Iazeolla:2008ix}
C.~Iazeolla and P.~Sundell, ``{A Fiber Approach to Harmonic Analysis of
  Unfolded Higher- Spin Field Equations},''
  \href{http://dx.doi.org/10.1088/1126-6708/2008/10/022}{{\em JHEP} {\bf 10}
  (2008)  022},
\href{http://arxiv.org/abs/0806.1942}{{\tt arXiv:0806.1942 [hep-th]}}.

\bibitem{Boulanger:2008kw}
N.~Boulanger, C.~Iazeolla, and P.~Sundell, ``{Unfolding Mixed-Symmetry Fields
  in AdS and the BMV Conjecture: II. Oscillator Realization},''
  \href{http://dx.doi.org/10.1088/1126-6708/2009/07/014}{{\em JHEP} {\bf 07}
  (2009)  014},
\href{http://arxiv.org/abs/0812.4438}{{\tt arXiv:0812.4438 [hep-th]}}.

\bibitem{Sezgin:2002rt}
E.~Sezgin and P.~Sundell, ``{Massless higher spins and holography},''
  \href{http://dx.doi.org/10.1016/S0550-3213(02)00739-3}{{\em Nucl. Phys.} {\bf
  B644} (2002)  303--370},
\href{http://arxiv.org/abs/hep-th/0205131}{{\tt arXiv:hep-th/0205131}}.

\bibitem{Klebanov:2002ja}
I.~R. Klebanov and A.~M. Polyakov, ``{AdS dual of the critical O(N) vector
  model},'' \href{http://dx.doi.org/10.1016/S0370-2693(02)02980-5}{{\em Phys.
  Lett.} {\bf B550} (2002)  213--219},
\href{http://arxiv.org/abs/hep-th/0210114}{{\tt arXiv:hep-th/0210114}}.

\bibitem{Giombi:2010vg}
S.~Giombi and X.~Yin, ``{Higher Spins in AdS and Twistorial Holography},''
  \href{http://dx.doi.org/10.1007/JHEP04(2011)086}{{\em JHEP} {\bf 1104} (2011)
   086},
\href{http://arxiv.org/abs/1004.3736}{{\tt arXiv:1004.3736 [hep-th]}}.

\bibitem{Aharony:2011jz}
O.~Aharony, G.~Gur-Ari, and R.~Yacoby, ``{$d=3$ Bosonic Vector Models Coupled
  to Chern-Simons Gauge Theories},''
  \href{http://dx.doi.org/10.1007/JHEP03(2012)037}{{\em JHEP} {\bf 1203} (2012)
   037}, \href{http://arxiv.org/abs/1110.4382}{{\tt arXiv:1110.4382 [hep-th]}}.

\bibitem{Sezgin:2003pt}
E.~Sezgin and P.~Sundell, ``{Holography in 4D (super) higher spin theories and
  a test via cubic scalar couplings},''
  \href{http://dx.doi.org/10.1088/1126-6708/2005/07/044}{{\em JHEP} {\bf 07}
  (2005)  044},
\href{http://arxiv.org/abs/hep-th/0305040}{{\tt arXiv:hep-th/0305040}}.

\bibitem{Giombi:2011ya}
S.~Giombi and X.~Yin, ``{On Higher Spin Gauge Theory and the Critical O(N)
  Model},'' \href{http://dx.doi.org/10.1103/PhysRevD.85.086005}{{\em Phys.Rev.}
  {\bf D85} (2012)  086005},
\href{http://arxiv.org/abs/1105.4011}{{\tt arXiv:1105.4011 [hep-th]}}.

\end{thebibliography}

\end{document}